\documentclass[prc,nofootinbib]{revtex4}

\usepackage{graphicx}
\usepackage[usenames,dvipsnames]{color}
\usepackage{amsmath,amssymb,bbold}
\usepackage{epsfig,rotate,latexsym}
\usepackage{subfig}
\usepackage{hyperref}
\usepackage{comment}
\usepackage{slashed}
\usepackage{float}
\begin{document}

\title {Thermoelectric transport coefficients of quark matter}
\author{Aman Abhisek$^{1}$}
\email{aman@prl.res.in}
\author{Arpan Das $^{2}$}
\email{arpan.das@ifj.edu.pl}
\author{Deepak Kumar$^{1,3}$}
\email{deepakk@prl.res.in}
\author{Hiranmaya Mishra$^{1}$}
\email{hm@prl.res.in}

\affiliation{$^{1}$ Theory Division, Physical Research Laboratory, Navrangpura, Ahmedabad 380 009, India}
\affiliation{$^{2}$Institute  of  Nuclear  Physics  Polish  Academy  of  Sciences,  PL-31-342  Krak\'ow,  Poland}
\affiliation{$^{3}$ Indian Institute of Technology Gandhinagar, Gandhinagar 382 355, Gujarat, India}

\date{\today} 

\def\be{\begin{equation}}
\def\ee{\end{equation}}
\def\bearr{\begin{eqnarray}}
\def\eearr{\end{eqnarray}}
\def\vec#1{{\bf {#1}}}
\def\bfm#1{\mbox{\boldmath $#1$}}
\def\hf{\frac{1}{2}}
\def\sl{\hspace{-0.15cm}/}
\def\omit#1{_{\!\rlap{$\scriptscriptstyle \backslash$}
{\scriptscriptstyle #1}}}
\def\vec#1{\mathchoice
        {\mbox{\boldmath $#1$}}
        {\mbox{\boldmath $#1$}}
        {\mbox{\boldmath $\scriptstyle #1$}}
        {\mbox{\boldmath $\scriptscriptstyle #1$}}
}

\begin{abstract}
A thermal gradient and/or a chemical potential gradient
in a conducting medium can lead to an electric field,
an effect known as thermoelectric effect or Seebeck effect.
In the context of heavy-ion collisions, we estimate the thermoelectric transport coefficients for quark matter
within the ambit of the Nambu-Jona Lasinio (NJL) model. We estimate the thermal conductivity, electrical conductivity, and the Seebeck coefficient of hot and dense quark matter.
These coefficients are calculated using the relativistic Boltzmann transport equation within relaxation time approximation.
The relaxation times for the quarks are estimated from the quark-quark and quark-antiquark scattering through
in-medium meson exchange within the NJL model.

\end{abstract}


\maketitle

\section{Introduction} 
Heavy-ion collision experiments conducted at particle accelerators allow us to study the properties 
of fundamental constituents of nature, such as quarks and gluons. Experiments at Relativistic Heavy 
Ion Collider (RHIC) and Large Hadron Collider (LHC) indicate the formation of such a deconfined medium 
of quarks and gluons. The partonic medium so produced, behaves like a strongly interacting liquid 
with a small value of shear viscosity ($\eta$) to entropy density ($s$) ratio $(\eta/s)$,  cools down with expansion,
undergoes a transition to the hadronic phase
and finally free streams to the detector. One of the successful descriptions of the bulk evolution of 
such strongly interacting matter has been through relativistic hydrodynamics.
Transport coefficients are important input parameters that enter in such a dissipative hydrodynamic 
description as well as in transport simulations that have been used to describe the evolution of such matter 
produced in a heavy-ion collision. 

 Hydrodynamic studies of the heavy-ion collisions suggest that the medium produced has a very small ratio of shear viscosity to entropy density ($\eta/s$)\cite{Heinz:2013th, Romatschke:2007mq, Kovtun:2004de}. It is amongst the smallest of known materials suggesting the quark-gluon plasma (QGP) formed is the most perfect fluid. The value of this ratio estimated from experiments is also found to be close to the conjectured  KSS bound on the value of $\eta/s$\cite{Kovtun:2004de}. 
Just like shear viscosity determines the response to transverse momentum gradients there are other transport coefficients such as bulk viscosity, electrical conductivity, etc. which determine the response of the system to other such perturbations. Bulk viscosity \cite{Dobado:2012zf, Sasaki:2008fg, Sasaki:2008um,Karsch:2007jc,Finazzo:2014cna,Jeon:1995zm} determines the response to bulk stresses. It scales with the conformal anomaly ($\frac{\epsilon - 3P}{T^4}$) 
and is expected to be large near the phase transition as inferred from lattice calculations \cite{Bazavov:2009zn,Bazavov:2010pg}. The effect of such a 
large bulk viscosity to entropy ratio have been investigated on the particle spectrum and flow coefficients \cite{Bozek:2009dw,Rose:2014fba}. 
Electrical conductivity ($\sigma_{el}$) \cite{Tuchin:2010gx, Tuchin:2010vs, Inghirami:2016iru,
 Das:2017qfi, Greif:2016skc, Greif:2014oia, Puglisi:2014pda, Puglisi:2014sha, Cassing:2013iz, 
Steinert:2013fza, Aarts:2014nba, Aarts:2007wj, Amato:2013naa, Gupta:2003zh, Ding:2010ga, Kaczmarek:2013dya, 
Qin:2013aaa, Marty:2013ita, FernandezFraile:2005ka} is also important as the heavy-ion collisions may be associated with large electromagnetic fields. The magnetic field produced in non-central collisions has been estimated to be of the order of $\sim m_\pi^2$ at RHIC energy scales \cite{Kharzeev:2007jp,Skokov:2009qp,Li:2016tel,Inghirami:2019mkc,Inghirami:2018ziv,Shokri:2017xxn,Shokri:2018qcu,Tabatabaee:2020efb}. Such magnetic fields are amongst the strongest magnetic fields produced in nature and can affect various properties of the strongly interacting medium. They may also lead to interesting CP-violating effects such as chiral magnetic effect etc \cite{Kharzeev:2012ph}.
In a conducting medium, the evolution of the magnetic field depends on the electrical conductivity. Electrical conductivity modifies the decay of the magnetic field substantially in comparison with the decay of the magnetic field in vacuum.  Hence the estimation of the electrical conductivity of the strongly interacting medium is important regarding the decay of the magnetic field produced at the initial stage of heavy ion collision.  These transport coefficients have been estimated in perturbative QCD and effective models\cite{Greif:2017byw,
 Prakash:1993bt, Wiranata:2012br, Chakraborty:2010fr, Khvorostukhin:2010aj, Plumari:2012ep, Gorenstein:2007mw,
 NoronhaHostler:2012ug, Tiwari:2011km, Ghosh:2013cba, Lang:2015nca, Ghosh:2014qba, Wiranata:2014kva, 
Wiranata:2012vv, NoronhaHostler:2008ju, Kadam:2014cua, Kadam:2014xka, Ghosh:2014yea, Rose:2017bjz, Wesp:2011yy,
 Greif:2014oia}. At finite baryon densities, the other transport coefficient that is relevant is the coefficient of
thermal conductivity and has been studied in \cite{Denicol:2012vq, Kapusta:2012zb} both in the hadronic matter as well as partonic matter.

In the present investigation, we focus on the thermoelectric response of the strongly interacting quark matter
 produced in a 
heavy-ion collision. It is well known from  condensed matter systems that
in a conducting medium, a temperature gradient can result in the generation of an electric current known as the Seebeck 
effect. Due to a temperature gradient, there is a non zero gradient of charge carriers
leading to the generation of an electric field. A measure of the electric field produced in such a conducting medium
due to a temperature gradient is the Seebeck coefficient which is defined as the ratio of the electric field to 
the temperature gradient in the limit of vanishing electric current. Seebeck effect has been extensively studied 
in condensed matter systems such as superconductors, quantum dots, high-temperature cuprates, 
superconductor-ferromagnetic tunnel junctions, low dimensional organic metals, 
\cite{seebconds1,seebconds2,seebconds3,seebconds4,seebconds5,seebconds6,seebconds7,seebconds8,seebconds9}. 
Such a phenomenon could also be present in the thermal medium created in heavy-ion collisions. 
It may further be noted that, in condensed matter systems  a temperature gradient
is sufficient for thermoelectric effect as there is only one type of dominant charge carriers in these systems.
In the strongly interacting medium produced in heavy-ion collisions, on the otherhand,
 both positive and negative charges contribute 
to transport phenomena. For vanishing baryon chemical potential (quark chemical potential) with equal numbers of 
particles and antiparticles there is no net thermoelectric effect. A finite baryon chemical potential 
(quark chemical potential) is required for the thermoelectric effect to be observed. The strongly interacting matter 
at finite baryon density can be produced in low energy heavy-ion collisions at finite, e.g. at FAIR and NICA. Along with 
the temperature gradient, we also consider a gradient in the baryon (quark) chemical potential to estimate the 
Seebeck coefficient of the partonic medium. The gradient in the chemical potential has effects similar to the 
temperature gradient. Using Gibbs Duhem relation for a static medium one can express gradient in the baryon (quark) chemical potential to a gradient in temperature. Effect of the chemical potential gradient significantly affects the thermoelectric coefficients as has been demonstrated in Ref.\cite{Das:2020beh} for hadronic system.

Seebeck effect in the hadronic matter has been investigated previously by some of us within the framework of the Hadron resonance gas model \cite{Bhatt:2018ncr,Das:2020beh}. However,
the Hadron resonance gas model can only describe the hadronic medium at chemical freezeout whereas 
one expects deconfined partonic medium at the early stages of the heavy-ion collisions. 
In this investigation, we estimate the thermoelectric behavior of the partonic medium within the 
framework of the NJL model. Seebeck coefficient has also been estimated for the partonic matter
including effects of a magnetic field within 
a relaxation time approximation in Ref.s\cite{Dey:2020sbm,Zhang:2020efz}.
However, this has been attempted with the relaxation time estimated within perturbative QCD which may be
valid only for asymptotically high temperatures. Further, it ought to be mentioned that, 
the vacuum structure of QCD remain nontrivial near the critical temperature region with nonvanishing values for 
the quark-antiquark condensates associated with chiral symmetry breaking as well as Polyakov loop 
condensates associated with the physics of statistical confinement
\cite{Singha:2017jmq,Abhishek:2017pkp,Singh:2018wps,Ratti:2005jh}. Indeed, within the ambit of the 
NJL model, it was shown that the
temperature dependence of viscosity coefficients exhibits interesting behavior of phase transition with the shear viscosity to entropy ratio showing a minimum while the coefficient of bulk viscosity showing a maximum at the phase transition \cite{Deb:2016myz,Singha:2017jmq,Abhishek:2017pkp}. The crucial reason for this behavior was the estimation of relaxation time using medium dependent masses for the quarks as well as the exchanged mesons which reveal nontrivial dependence before and after the transition temperature. This motivates us to investigate the behavior of thermoelectric transport coefficients within the NJL model which takes into account
the medium dependence of quark and meson masses.
This model has been used to study different transport properties of quark matter at high temperatures \cite{Marty:2013ita, Deb:2016myz, Rehberg:1995kh, Sasaki:2008um} and high densities\cite{Tsue:2012jz, Tatsumi2011, Menezes:2008qt, Chatterjee:2011ry, Mandal:2009uk, Mandal:2012fq, Mandal:2016dzg, Coppola:2017edn}. 
        
We organize the paper in the following manner. In Sec.~\eqref{formalism}, we discuss the Boltzmann equation within relaxation time approximation to have the expressions for the different thermoelectric transport coefficients when the quasi-particles have a medium dependent masses.
In Sec.~\eqref{NJLmodel} we discuss thermodynamics and estimation of relaxation time within the two flavor NJL model. In Sec.~\eqref{results} we present the results of different transport coefficients. Finally, we give a possible outlook of the present investigation and conclude in Sec.~\eqref{conclusion}. 

%%%%%%%%%%%%%%%%%%%%%%%%%%%%%%%%%%%%%%%%%%%%
%%%%%%%%%%%%%%%%%%%%%%%%%%%%%%%%%%%%%%%%%%%%

\section{ Boltzmann equation in relaxation time approximation and transport coefficients}
\label{formalism}

Within a quasiparticle approximation, a kinetic theory treatment for the calculation of transport
coefficient can be a reasonable approximation that we shall be following similar to 
that in Refs. \cite{Sasaki:2008fg,Sasaki:2008um,Chakraborty:2010fr,Khvorostukhin:2010aj,Khvorostukhin:2012kw,Zhuang:1995uf}. The plasma can be described by a phase space density for each species of particle. Near equilibrium, the distribution function can be expanded about a local equilibrium distribution function for the quarks as,
$$
f(\vec{x},\vec{p},t) =f^{(0)}(\vec{x},\vec{p})+\delta f(\vec{x},\vec{p}, t),
$$
where the local equilibrium distribution function $f^{(0)}$ is given as
\be
f^{(0)}(\vec{x},\vec{p})=\left[\exp\left(\beta(\vec x)\left(u_\nu p^\nu\mp\mu(\vec{x})\right)\right)+1\right]^{-1}.
\label{f0}
\ee
Here, $u^\mu=\gamma_u(1,\vec u)$,  is the flow four-velocity, where,  $\gamma_u=(1-\vec u^2)^{1/2}$; $\mu$ is the chemical
potential associated with a conserved charge. Here $\mu$ denotes the quark chemical potential  and $\beta=1/T$ is the inverse of temperature. Further, $p^{\mu}=(E,\vec{p})$ is the particle four momenta, single particle energy $E=\sqrt{p^2+M^2}$ with $p=|\vec{p}|$. $M$ is the mass of the particle which in general is medium dependent. The departure from the equilibrium is 
described by the Boltzmann equation,
\be
\frac{df_a(\vec{x},\vec{p},t)}{dt}=\frac{\partial f_a}{\partial t}+\frac{dx^i}{dt}\frac{\partial f_a}{\partial x^i}
+\frac{dp^i}{dt}\frac{\partial f_a}{\partial p^i}=C^a[f],
\label{boltzeq}
\ee
where we have introduced the species index `$a$' on the distribution function. The right-hand side is the collision term which we shall discuss later. The left-hand side of the Boltzmann equation involves the trajectory $\vec{x}(t)$ and the momentum $\vec{p}(t)$. This trajectory, in general, not a straight line as the particle is moving in a mean-field, which, in general, can be space time-dependent. The velocity of the particle `$a$' is given by
$$\frac{d x^i}{d t}=\frac{\partial E_a}{\partial p_a^i}=\frac{p_a^i}{E_a}=v_a^i.$$

Next, the time derivative of momentum , the force, in presence of an electric field $(\vec{\mathcal{E}})$, magnetic field $(\vec{B})$ and a mean field dependent mass can be written as
$$\frac{dp^i}{dt}=-\frac{\partial E_a}{\partial x^i}+q_a(\mathcal{E}^i+\epsilon^{ijk}v_jB_k).$$

 The time derivatives of $\vec x$ and $\vec p$ can be substituted on the left-hand side of the Boltzmann equation 
Eq.(\ref{boltzeq}) and the same reduces to
\be
\frac{\partial f_a}{\partial t}+ v^i \frac{\partial f_a}{\partial x^i}
+\frac{\partial f_a}{\partial p^i}\left(-\frac{M_a}{E_a}\frac{\partial M_a}{\partial x^i}+
q_a(\mathcal{E}^i+\epsilon^{ijk}v_jB_k)\right)=C^a[f].
\label{boltzeq1}
\ee

For the collision term on the right-hand side, we shall be
limiting ourselves to $2\rightarrow 2$ scatterings only. In the relaxation time approximation the collision term for species $a$, all the distribution functions are given by the equilibrium distribution function except the distribution function for particle $a$. The collision term, to first order in the deviation from the equilibrium function,
will then be proportional to $\delta f_a$, given the fact that $C^a[f^{(0)}]=0$ by the principle of local detailed balance.
In that case, the collision term is given by
\be
C[f]=-\frac{\delta f_a}{\tau_a},
\ee
 where, $\tau_a$, the relaxation time for particle `$a$'. In general relaxation time is a function of energy. We shall discuss more about it in the subsequent subsection where we estimate it within the NJL model. Returning back to the left-hand side of Eq.(\ref{boltzeq1}), we keep up to the first order in gradients in space-time. The left-hand side of the Boltzmann equation Eq.(\ref{boltzeq1}), is explicitly small because of the gradients and we, therefore, may
replace $f_a$ by $f^{(0)}_a$. While the spatial derivative of the distribution function is given by,
\be
\frac{\partial f^{(0)}_a}{\partial x^i}=-f^{(0)}_a(1-f^{(0)}_a)\partial_i(\beta E_a-\beta\mu_a)=
-f^{(0)}_a(1-f^{(0)}_a)\left(-\frac{E_a}{T^2}\partial_i T+\beta\frac{M_a}{E_a}\frac{\partial M_a}{\partial x^i}
-\partial_i(\beta\mu_a)\right),
\label{delxf}
\ee
here $\mu_a = b_a\mu$, $b_a$ being the quark number, i.e. $b_a=1$ for quarks and $b_a=-1$ for antiquarks. The momentum derivative of the equilibrium distribution function is given by,
\be
\frac{\partial f^{(0)}_a}{\partial p^i}=-\frac{1}{T}f^{(0)}_a(1-f^{(0)}_a)v_a^i.
\label{delpf}
\ee

Substituting Eqs.\eqref{delpf} and \eqref{delxf} in the Boltzmann equation Eq.(\ref{boltzeq1}) for the static case (where the distribution function is not an explicit function of time) in the absence of magnetic field we have
\be
-f_a^{(0)}(1-f_a^{(0)})\left[v_a^i\left(-\frac{1}{T^2}\partial_iT E_a-\partial_i (\beta\mu_a)\right)
+q_a\beta  v_a^i\mathcal{E}^i\right]=-\frac{\delta f_a}{\tau_a}.
\label{boltztau}
\ee

The spatial gradients of temperature and chemical potential can be related using
momentum conservation in the system and Gibbs Duhem relation. Momentum conservation in a steady-state leads to  $\partial_i P=0$ ( $P$, being the pressure)\cite{Gavin:1985ph}. Using Gibbs Duhem relation, the pressure gradient can be written as, with the enthalpy $\omega=\epsilon+P$,
\begin{equation}
\partial _i P= \frac{\omega}{T}\partial_i T+T n_q\partial_i(\mu/T)
\end{equation}
which vanishes in steady-state. $n_q$ denotes the net quark number density and $\epsilon$ is the energy density. The above equation relates the spatial gradient of temperature to the spatial gradients in chemical potential as,
\begin{equation}
\partial_i\mu=\left(\mu-\frac{\omega}{n_q}\right)\frac{\partial_i T}{T}.
\label{tmuder}
\end{equation}

Using Eq.\eqref{tmuder} and Eq.\eqref{boltztau}, $\delta f_a$,
the deviation of the distribution function is given as,
\be
\delta f_a=\frac{\tau_a f_a^0(1-f_a^0)}{T}\left[q_a\vec{v}_a\cdot\vec{\mathcal{E}}-\left(E_a-b_a\frac{\omega}{n_q}\right)\frac{\vec v_a\cdot\vec{\nabla} T}{T}\right].
\label{delfa}
\ee

 The nonequilibrium part of the distribution function gives rise to transport coefficients. The electric current is now given as,
\bearr
 \vec J& = &\sum_a g_a\int \frac{d^3p_a}{(2\pi)^3}~q_a\vec{v}_a~\delta f_a\nonumber\\
&=& \sum_a\frac{g_aq_a^2}{3T}\int \frac {d^3 p_a}{(2\pi)^3}~v_a^2\tau_a f_a^0(1-f_a^0)~ \vec{\mathcal{E}}\nonumber\\
&-& \sum_a\frac{g_aq_a}{3T^2}\int \frac{d^3 p_a}{(2\pi)^3}~\tau_a\left(E_a-b_a\frac{\omega}{n_q}\right)f_a^0(1-f_a^0)
v_a^2~\vec{\nabla}T.
\label{jelec}
\eearr

In Eq.\eqref{jelec} we have used $ v_a^iv_a^j=\frac{1}{3}v_a^2\delta^{ij}$ as because the integrand only depends on the magnitude of momenta. Further, the sum is over all 
flavors including antiparticles. The degeneracy factor $g_a=6$ corresponding to color and spin degrees of freedom.
$b_a$ is the quark number i.e. $b_a=\pm 1$ for quarks and antiquarks respectively.

Next, we write down the heat current $\vec{\mathcal{I}}$ associated with the conserved quark number.
For a relativistic system, thermal current arises corresponding to a conserved particle number. 
 The thermal conduction due to quarks arises when there is energy flow relative to enthalpy~\cite{Gavin:1985ph}.
Therefore the heat current is defined  as \cite{Gavin:1985ph}, 
\be
\mathcal{I}^i=\sum_aT_a^{0i}-\frac{\omega}{n_q}\sum_ab_a J_{qa}^i.
\label{hcurr}
\ee
Here, $n_q$ is the net quark number density.
The energy flux is given by $T^{0i}$, the spatio-temporal component of energy-momentum tensor ($T^{\mu\nu}$)\cite{Gavin:1985ph},
\be
T^{0i}_a=g_a\int\frac{d^3 p_a}{(2\pi)^3}p_a^if_a.
\label{EMtensor}
\ee
While, quark current is given $\vec J_q$ is given by
\be
 J^i_{qa}=g_a\int\frac{d^3 p_a}{(2\pi)^3}\frac{p^i_a}{E_a}f^{}_a b_a,
\label{qcurr}
\ee

Clearly, the contribution to the energy flux and quark current vanishes arising from the equilibrium distribution function $f_a^{(0)}$ due to symmetry consideration and it is only the nonequilibrium part $\delta f_a$ that contribute to the energy flux and quark current in Eqs.\eqref{EMtensor} and Eq.\eqref{qcurr} respectively. 
Substituting the expression for $\delta f_a$ from Eq.\eqref{delfa} in Eq.\eqref{hcurr}, the heat current $\vec{\mathcal{I}}$ is given as,

\be
\vec{\mathcal{I}}=\sum_a \frac{g_a}{3T}\int \frac{d^3 p_a}{(2\pi)^3}f_a^0(1-f_a^0) v_a^2\tau_a\left[q_a\left(E_a-b_a\frac{\omega}{n_q}\right)\vec{\mathcal{E}}
-\left(E_a-b_a\frac{\omega}{n_q}\right)^2\frac{\vec{\nabla}T}{T}\right]
\label{hcurr1}
\ee

The Seebeck coefficient $ S$ is defined by setting the electric current $\vec J=0$ in Eq.(\ref{jelec}) so that
the electric field becomes proportional to the temperature gradient i.e.
\be
\vec{\mathcal{E}}=S\vec{\nabla}T.
\ee
Therefore the Seebeck coefficient for the quark matter in the presence of a gradient in temperature and chemical potential can be expressed as,
\be
S=\frac{\sum_a\frac{g_aq_a}{3T}\int \frac {d^3 p_a}{(2\pi)^3}\tau_a  v^2
\left(E_a-b_a\frac{\omega}{n_q}\right)f_a^{(0)}(1-f_a^{(0)})}{T
\sum_a\frac{g_a}{3T}q_a^2\int\frac{d^3 p_a}{(2\pi)^3} v^2 \tau_a f_a^{(0)}(1-f_a^{(0)})}
\label{seeq}
\ee
The denominator of the Seebeck coefficient in the above may be identified as $T \sigma_{el}$, where the 
electrical conductivity $\sigma_{el}$ is given by\cite{Puglisi:2014sha,Kadam:2017iaz},
\be
\sigma_{el}= 
\sum_a\frac{g_a}{3T}q_a^2\int\frac{d^3 p_a}{(2\pi)^3}
\left(\frac{p_a}{E_a}\right)^2 
\tau_a f_a^{(0)}(1-f_a^{(0)})
\label{econd}
\ee
% \end{document}
which may be identified from Eq.\eqref{jelec}. Let us note that, while the denominator of the
Seebeck coefficient is positive definite, the numerator is not so as it is linearly dependent on the electric charge of the
species as well as on the difference $(E_a-b_a\frac{\omega}{n_q})$. This makes the Seebeck coefficient not always positive
definite. This is also observed in different condensed matter systems \cite{Zhou:2020}. 

In terms of the electrical conductivity and the Seebeck coefficient, the electric current Eq.(\ref{jelec})
can be written as
\be
\vec J=\sigma_{el}\vec{\mathcal{E}}-\sigma_{el}S\vec\nabla T.
\label{equnew19}
\ee

In a similar manner, the heat current as given in Eq.(\ref{hcurr1}) can be written as,
\be
\vec{\mathcal{I}}=T\sigma_{el}S\vec{\mathcal{E}}-\kappa_0 \vec\nabla T,
\label{equnew20}
\ee
where, $\kappa_0$, the thermal conductivity can be written as\cite{Gavin:1985ph}
\be
\kappa_0=\sum_a\frac{g_a}{3T^2}\int \frac{d^3 p_a}{(2\pi)^3}
\tau_a\left(\frac{p_a}{E_a}\right)^2
\left(E_a-b_a\frac{\omega}{n_q}\right)^2f_a^{(0)}(1-f_a^{(0)}).
\label{tcond}
\ee

Using Eqs.\eqref{equnew19} and \eqref{equnew20}, we can express the heat current ($\vec{\mathcal{I}}$) in terms of electric current ($\vec{J}$) in the following way,
\be
\vec{\mathcal{I}}=T S \vec{J}-\left(\kappa_{0}-T \sigma_{el} S^{2}\right) \vec{\nabla} T.
\label{equnew22}
\ee
From Eq.\eqref{equnew22} we can identify the Peltier coefficient ($\Pi$) and thermal conductivity ( $k$ ) in the presence of nonvanishing Seebeck coefficient as,

\be
\Pi=T S, ~~
\kappa=\kappa_{0}- T \sigma_{el} S^{2}.
\label{equnew23}
\ee

Note that the relation between the Peltier coefficient ($\Pi$) and the Seebeck coefficient as given in Eq.\eqref{equnew23} can be considered
as the consistency relation. Also, note that the
thermal conductivity in the absence of any thermoelectric effect as given in Eq.\eqref{tcond} matches with the expression of the thermal conductivity as reported in \cite{Gavin:1985ph}.
The Seebeck coefficient ($S$), thermal conductivity ($\kappa_0$), and the electrical conductivity ($\sigma_{el}$) depend upon, the estimation of the
relaxation time as well as the quark masses that goes into the distribution functions through the single-particle energies
and are medium dependent. We estimate these quantities in the Nambu-Jona-Lasinio model which is described in the next section. 

%%%%%%%%%%%%%%%%%%%%%%%%%%%%%%%%%%%%%%%%%%%%
%%%%%%%%%%%%%%%%%%%%%%%%%%%%%%%%%%%%%%%%%%%%%

\section{Estimation of relaxation time in NJL model}
\label{NJLmodel}
We model the partonic medium using the two flavor Nambu-Jona-Lasinio (NJL) model and estimate
the thermodynamic quantities, the quasi particle masses in the medium and the relaxation time. The two flavour NJL model with $u$ and $d$ quark, can be described by the following  Lagrangian~\cite{Buballa:2003qv},
\be
\mathcal {L}=\bar\psi(i\slashed{\partial}-m_q)\psi+G\left[(\bar\psi\psi)^2+(\bar\psi i\gamma^5\vec\tau\psi)^2\right].
\label{njllag}
\ee
Here, $\psi$ is the doublet of $u$ and $d$ quarks;
$m_q$ is the current quark mass matrix which is diagonal with elements $m_u$ and $m_d$ and we take them to be same as $m_0$
assuming isospin symmetry;
$\vec \tau$ are the Pauli matrices in the flavor space; $G$ is the scalar coupling. NJL model is a QCD inspired effective model which incorporates various aspects of the chiral symmetry of QCD. 
The NJL model Lagrangian as given in Eq.~\eqref{njllag} is symmetric under the chiral symmetry group 
$SU(2)_V\times SU(2)_A\times U(1)_V$. The thermodynamic quantities e.g., pressure ($P$), energy density ($\epsilon$)
 and the number density $(n_q)$ can be obtained once we know the thermodynamic potential of the NJL model. 
In a grand canonical ensemble, the thermodynamic potential ($\Omega$) or equivalently the pressure ($P$) can be expressed as,  
\be
-P=\Omega(\beta,\mu)=\frac{(M-m_0)^2}{4G}-\frac{2N_cN_f}{(2\pi)^3\beta}\int d\vec k\left[\log(1+e^{-\beta(E-\mu)})
+\log(1+e^{-\beta(E+\mu)})\right]-\frac{2N_c N_f}{(2\pi)^3}\int d\vec k \sqrt{\vec k^2+M^2},
\label{pres}
\ee
%$\epsilon_{vac}$ is the Dirac sea contribution given as,
%\be
%\epsilon_{vac}=\frac{12}{(2\pi)^3}\int_{|\vec k|\leq\Lambda} d^3 k \left[\sqrt{k^2+M^2}-\sqrt{ k^2+m_0^2}\right].
%\ee
%$\epsilon_{vac}$ is written in such a way that for the perturbative vacuum and zero temperature and vanishing density, the energy density of the perturbative vacuum vanishes.
% vacuum the pressure is zero for the perturbative vacuum.
here in the intergrals $d\vec k$ denotes $d^3k$. In the above, $N_c=3$ is the number of colors and $N_f=2$ is the number of flavors, $E=\sqrt{\vec k^2+M^2}$
is the single particle energy with `constituent' quark mass M which satisfies the self consistent gap equation
\be
M=m_0+\frac{2N_cN_f}{(2\pi)^3}\int d\vec k \frac{M}{\sqrt{ k^2+M^2}} (1-f^{(0)}-\bar f^{(0)}).
\label{gapeq}
\ee
In the above equations $f^{(0)}=(1+\exp(\beta\omega_-))^{-1}$ and $\bar f^{(0)}=(1+\exp(\beta\omega_+))^{-1}$  are the 
equilibrium distribution functions for quarks and antiquarks respectively and we have written 
 $\omega_{\pm}( k)=E( \vec k)\pm\mu$ with $k\equiv|\vec{k}|$. The energy density $\epsilon$
is given by,
\be
\epsilon=-\frac{2 N_cN_f}{(2\pi)^3}\int d\vec k E(k)(1-f^{(0)}-\bar f^{(0)})+\frac{(M-m_0)^2}{4G},
\ee
so that enthalpy $\omega=\epsilon+P$ is also defined once the solution to the mass gap equation Eq.(\ref{gapeq}) is known.
In these calculations, we have taken a three momentum cutoff $\Lambda$ for the for calculations of integrals not involving 
the Fermi distribution functions.
The net number density of quarks $n_q$ is given as
\be
n_q=\frac{2 N_c N_f}{(2\pi)^3}\int d\vec{k} (f^{(0)}-\bar f^{(0)}).
\ee
This completes the discussion on the all the bulk thermodynamic quantities defined for NJL model which enters in the definitions
for Seebeck coefficient, electrical conductivity and thermal conductivity.

Next we discuss  the estimation of relaxation time and as mentioned earlier we consider two particle 
scattering processes only. For a process $a+b\rightarrow c+d$, the relaxation time for the particle $a$ i.e. $\tau_a(E_a)$ is given by~\cite{Deb:2016myz},
\be
\tau_a^{-1}(E_a)\equiv\tilde{\omega}(E_a)=\frac{1}{2 E_a}\sum_b\int d\vec\pi_b W_{ab}f_b^{(0)}(E_b),
\label{relaxa}
\ee
where, the summation is over all species other than the particle $``a"$. Further, in Eq.(\ref{relaxa}), we have
introduced the notation $d\vec\pi_i=\frac{d^3 p_i}{(2\pi)^3 2 E_i}$ and  $W_{ab}$ is the
dimensionless transition rate for the processes with $a,b$ as the initial states. $W_{ab}$ which is Lorentz invariant and a function of the Mandelstam variable ($s$) can be given by,
\be
W_{ab}(s)=\frac{1}{1+\delta_{ab}}\int d\vec\pi_cd\vec\pi_d (2\pi)^4\delta(p_a+p_b-p_c-p_d)|\mathcal{M}|_{ab\rightarrow cd}^2
(1-f_c^{(0)}(p_c))(1-f_d^{(0)}(p_d)).
\ee
In the expression of $W_{ab}$ the Pauli blocking factors have been considered. The quantity $W_{a b}$ can be related to the cross sections of various scattering processes. In the present case within the NJL model, the quark-quark, quark-antiquark and antiquark-antiquark scattering cross sections are calculated to order $1/N_c$ which occur through the $\pi$ and $\sigma$ meson exchanges in ``$s$'' and ``$t$'' channels. The meson propagators that enters into the scattering amplitude is calculated within the random phase approximation and includes their masses and the widths. The mass of the meson  is estimated from the pole of the meson propagator at vanishing three momentum i.e.,
\be
1-2G~\text{Re}\Pi_{\tilde m}(M_{\tilde m },0)=0.
\label{mesonmass}
\ee 
where $\tilde{m}$ denotes $\sigma, \pi$ for scalar and pseudoscalar channel mesons, respectively. 
Polarization function in the corresponding mesonic channel is expressed as $\Pi_{\tilde{m}}$.
The explicit expressions for $\text{Re}\Pi_{\tilde{m}}$ and the imaginary part $\text{Im}\Pi_{\tilde{m}}$ is given in Ref.\cite{Deb:2016myz} and we do not repeat here.

While, the relaxation time is energy dependent, one can also define an energy independent mean relaxation time
by taking a thermal average as, 
\be
\bar\omega_a\equiv\bar \tau^{-1}_a=\frac{1}{n_a}\int \frac{d^3 p_a}{(2\pi)^3} f^{(0)}_a(E_a)\tilde{\omega}_a(E_a)\equiv\sum_b n_b
\bar W_{ab},
\label{bartau}
\ee
to get an estimate of the average relaxation time. In the above equation,
the sum is over all the particles other than `a';  $$n_a=\int \frac{d^3 p_a}{(2\pi)^3} f^{(0)}_a(E_a),$$ is the number density of the species ``$a$'' apart from the degeneracy factor.
Here, $\bar W_{ab}$ is the thermal-averaged transition rate given as
\be
\bar W_{ab}=\frac{1}{n_an_b}\int d\vec\pi_a d\vec\pi_b f(E_a)f(E_b) W_{ab}.
\label{barw}
\ee

 For the case of two flavors, there are 12 different processes but the corresponding matrix elements can be 
related using i-spin symmetry, charge conjugation and crossing symmetries with only two independent matrix 
elements. We have chosen them, as in Refs.\cite{Zhuang:1995uf,Deb:2016myz}, to be the processes $u\bar u\rightarrow u\bar u$ and $u\bar d\rightarrow u\bar d$. 
 The explicit expressions for the matrix elements are given in Refs.\cite{Zhuang:1995uf,Deb:2016myz}. 
In the meson propagators we have kept both the mass and the width of the meson resonances which are medium dependent.
It is important to mention that while the matrix elements of different scattering processes are related, 
the thermal-averaged rates are not. This is because the thermal averaged rates involve also the 
thermal distribution functions for the initial states along with the Pauli blocking factors for the final states.

%%%%%%%%%%%%%%%%%%%%%%%%%%%%%%%%%%%%%%%%%%
%%%%%%%%%%%%%%%%%%%%%%%%%%%%%%%%%%%%%%%%%%

\section{Results}
\label{results}
The two flavor NJL model as given in Eq.(\ref{njllag}) has three parameters, the four fermions coupling $G$, the three momenta cut off ($\Lambda$) to regularize the momentum integral in vacuum and the current quark mass $m_0$. These values are adjusted to fit the physical values of the pion mass ($m_{\pi}$=135 MeV), the pion decay constant
($f_{\pi}$=94 MeV) and the value of the quark condensate in vacuum, $\langle \bar u u\rangle=\langle\bar d d\rangle=(-241 ~\text{MeV})^3$ .
We have considered here the value of the parameters as $m_0=5.6$ MeV, $\Lambda=587.9$ MeV and $G\Lambda^2=2.44$ \cite{Buballa:2003qv}. 
This leads to the constituent quark mass for $u$ and $d$ type quarks, $M=397$ MeV in vacuum ($T=0,\mu=0$).

\begin{figure}
    \begin{minipage}{.485\textwidth}
        \centering
        \includegraphics[width=1.1\textwidth]{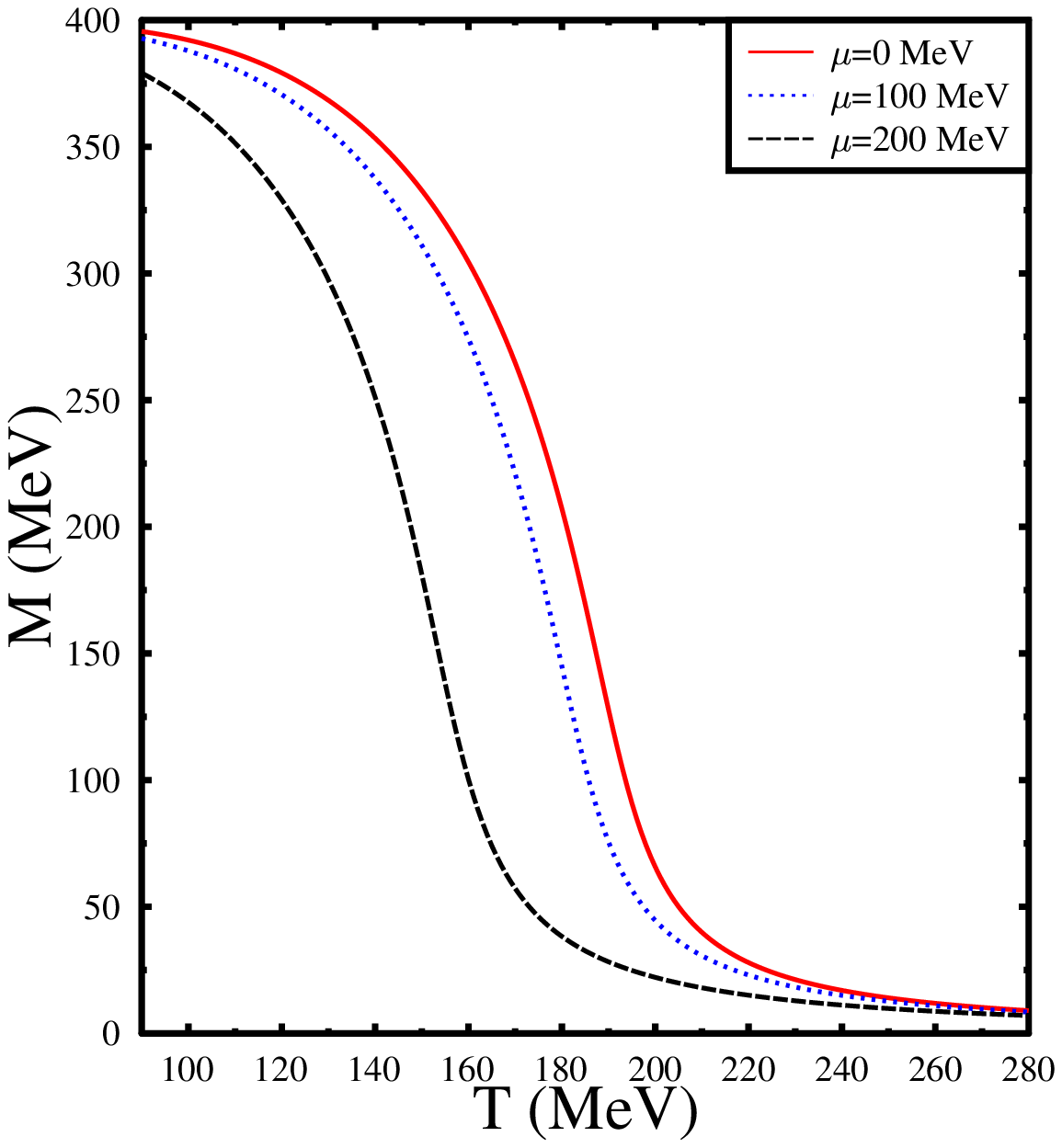}
    \end{minipage}
    \begin{minipage}{.485\textwidth}
        \centering
        \includegraphics[width=1.1\textwidth]{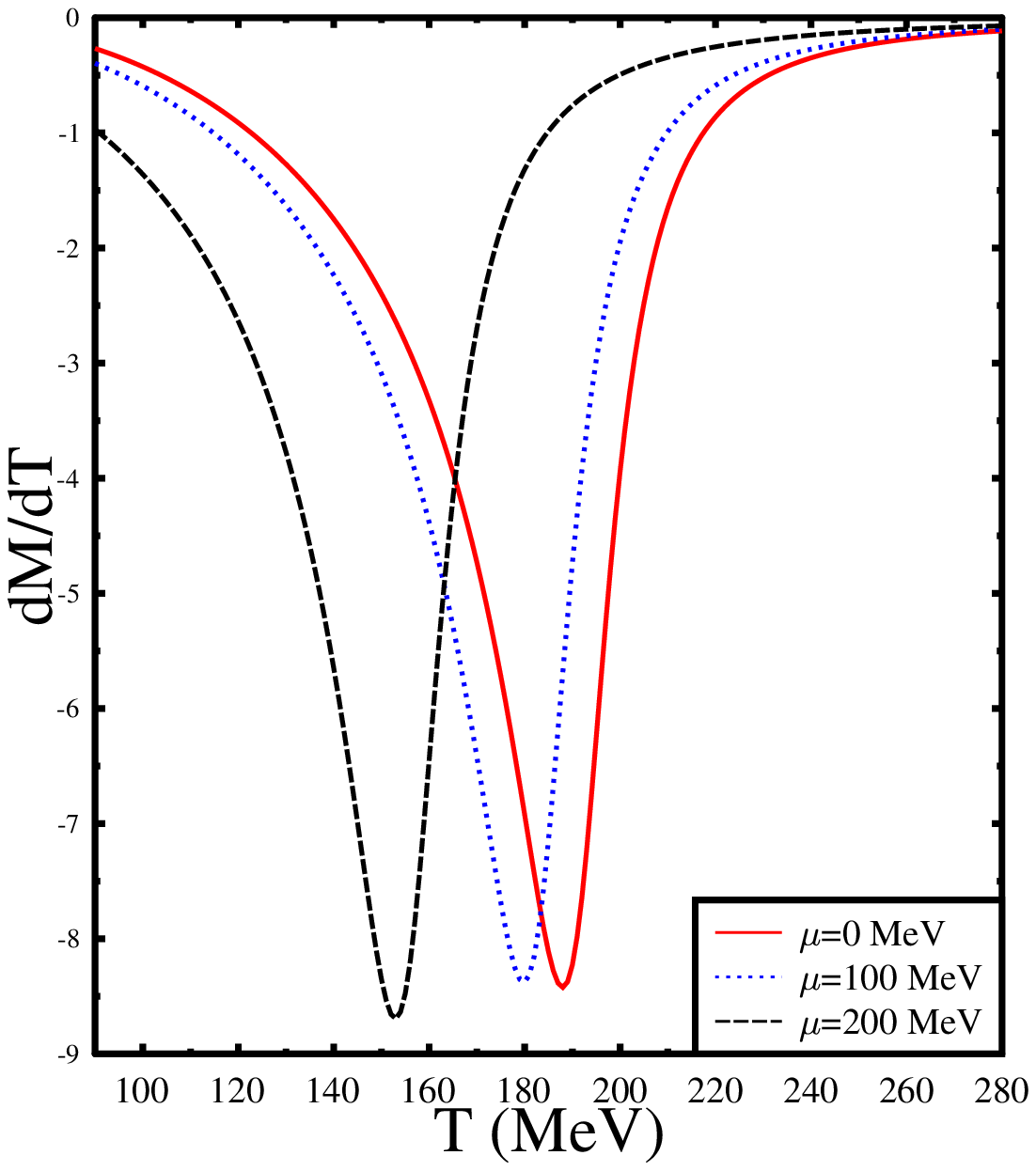}
    \end{minipage}  
    \caption{Left plot: temperature dependence of the masses of constituent quarks ($M$) for different chemical potentials. Right plot: variation of $dM/dT$ with temperature for different chemical potentials. The nonmonotonic variation of $dM/dT$ with a peak structure indicate the pseudo critical temperature for the chiral transition. Note that for the NJL model parameter set and the range of temperature and chemical potential considered here the chiral transition is a smooth crossover.}
 \label{fig:1}
\end{figure}

 To analyze the variation of different transport coefficients with temperature and quark chemical potential, we have first plotted in the left plot of Fig.~\eqref{fig:1}, the constituent quark masses ($M$) as a function of temperature ($T$) for different values of the quark chemical potential ($\mu$). The constituent quark mass ($M$) results as a solution to the gap equation, Eq.(\ref{gapeq}). Constituent quark masses for $u$ and $d$ quarks are the same and they are related to the quark-antiquark condensate $\langle\bar\psi\psi\rangle$. In the right plot of Fig.~\eqref{fig:1}, we have plotted $dM/dT$ with temperature for different values of the chemical potential. For the range of temperature and chemical potential considered here the chiral transition is a smooth crossover.  The chiral crossover temperature may be defined by the position of the peak in the variation of $dM/dT$ with temperature. For $\mu=$ 0, 100 and 200 MeV, the corresponding chiral crossover temperatures turns out to be $\sim $ 188 MeV, 180 MeV and 153 MeV respectively. It is expected that with an increase in chemical potential the crossover temperature decreases. Note that we have considered here the values of the chemical potential which are lower than the chemical potential corresponding to the speculated critical endpoint of the quark-hadron phase transition in the QCD phase diagram. 

\begin{figure}
        \centering
        \includegraphics[width=0.65\textwidth]{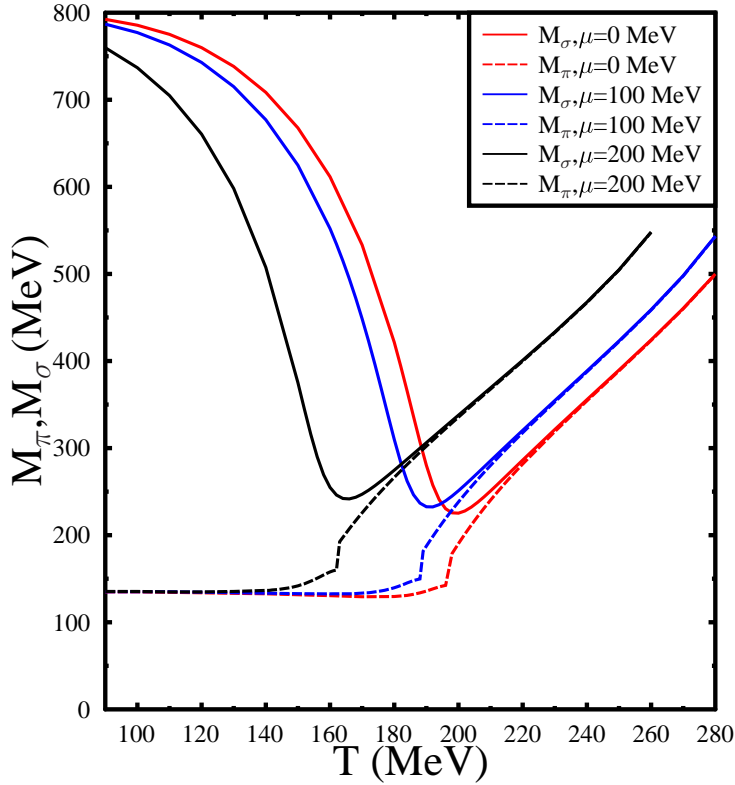}
    \caption{Variation of $\sigma$ and $\pi$ meson masses with temperature for different values of the chemical potentials. The solid lines correspond to $M_{\sigma}$ while the dashed lines correspond to pion masses, $M_\pi$.}
 \label{fig:2}
\end{figure}

In Fig.~\eqref{fig:2} we have plotted the meson masses $M_\pi$ and $M_\sigma$ as a function of temperature for different values of chemical potential as solutions of Eq.(\ref{mesonmass}).
Note that pions are pseudo-Goldstone modes, therefore in the chiral symmetry broken phase pion mass varies weakly.
But $M_\sigma$ decreases rapidly near the crossover temperature. At higher temperatures, $M_\pi$ and $M_\sigma$, being chiral partners, become approximately degenerate and increase with temperature. Further one can define a characteristic temperature, the ``Mott temperature" ($T_M$) where the pion mass becomes twice that of quark mass i.e. at Mott temperature $M_\pi(T_M)=2M(T_M)$. The Mott temperatures for $\mu$=0, 100 and 200 MeV turns out to be $\sim$ $198$ MeV, 192 MeV and 166 MeV respectively. As we shall see later it is the Mott temperature that becomes relevant while estimating the relaxation times of the quarks using thermal scattering rates of the quarks through meson exchange.

\begin{figure}
        \centering
        \includegraphics[width=0.65\textwidth]{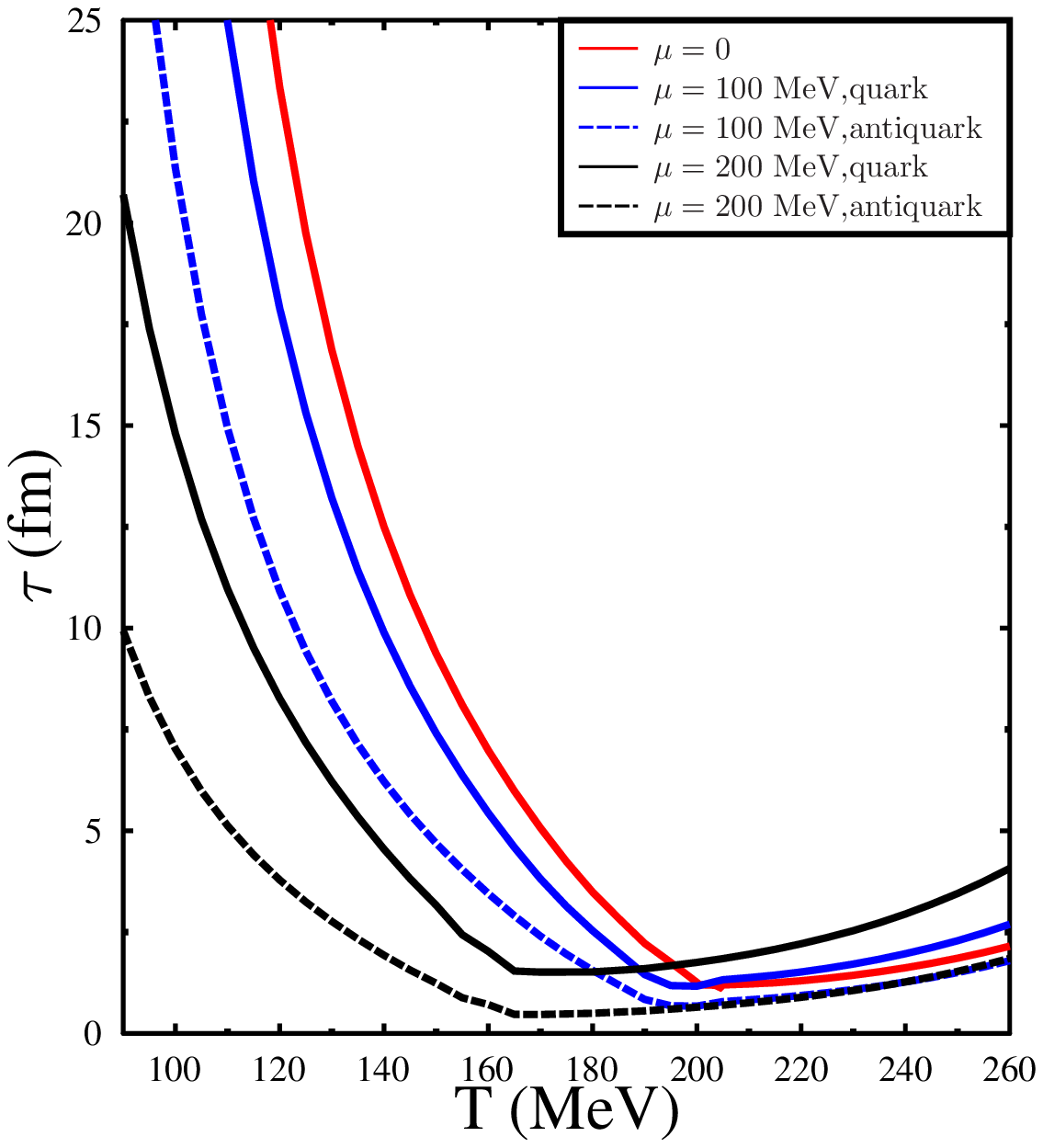}
    \caption{Variation of thermal averaged relaxation times for quarks and antiquarks  with temperature for different chemical potentials.
    Solid lines correspond to  the relaxation time for
quarks while the dotted lines correspond to relaxation time for antiquarks. For $\mu=0$ the thermal averaged relaxation times for the quarks and antiquarks are same. Difference between the relaxation times of quarks and antiquarks appears only at finite chemical potential.}
 \label{fig:3}
\end{figure}

In Fig.~\eqref{fig:3}, we show the variation of average relaxation time as defined in Eq.(\ref{bartau}), 
for quarks (solid lines) and antiquarks(dashed lines) with temperature for different chemical potentials. Let us 
note that the relaxation time of given particle '$a$', as shown in Eq.(\ref{bartau}),
depends both on the scattering rates $\bar W_{ab}$ as well as on the number 
density $n_b$ of the particles other than '$a$' in the initial state i.e. number density of scatterers.
It turns out that, for the scattering processes considered here,
the process $u\bar{d}\rightarrow u\bar{d}$ \cite{Deb:2016myz}, through charged pion exchange in the s-channel
gives the largest contribution  to the scattering rate $\bar W_{ab}$ as compared to other channels. As mentioned earlier,by crossing symmetry arguments,
 this also means that the $ud\rightarrow u d$ scattering rate also contribute dominantly to the thermally averaged scattering
rate.

Let us discuss first the behaviour of the relaxation time below the Mott temperature $T_M$.
Below $T_M$, the average scattering rate is suppresed mostly due to
thermal distribution with large constituent quark masses apart from the suppression from the sigma meson propagators 
with large $M_\sigma$
in the scattering amplitudes. As one approaches $T_M$ from lower temperature, the scattering rates become larger as
the constituent quark mass decreases leading to a decrease of the relaxation time for quarks as well as antiquarks.
Further, as the chemical potential increases, the densities of antiquarks gets suppressed leading to larger
relaxation time for quarks compared to antiquarks. This is what is observed for the behaviour
of relaxation time as a function of $T$ and $\mu$ in Fig. ~\eqref{fig:3} below thw Mott temperature.

Above $T_M$, the meson propagator develop a pole in the s-channel
leading to an enhancement of the scatterring rate.
However, at large temperature beyond $T_M$, there will be a suppression due to the 
large meson masses which increase with temperature. This results in a maximum scattering rate at $T_M$ or a minimum in the 
average relaxation time as generically seen in Fig.~\eqref{fig:3}. 

%At finite chemical potential, the number density of quarks get enhanced while that of antiquarks get suppressed. Due to this, the dominant contribution $u\bar d\rightarrow u\bar d$
%to the scatterring rate $\bar\omega_a$ gets suppressed for quarks and enhanced for antiquarks as comapred to
%the case of $\mu=0$. 

\begin{figure}
    \begin{minipage}{.485\textwidth}
        \centering
        \includegraphics[width=1.1\textwidth]{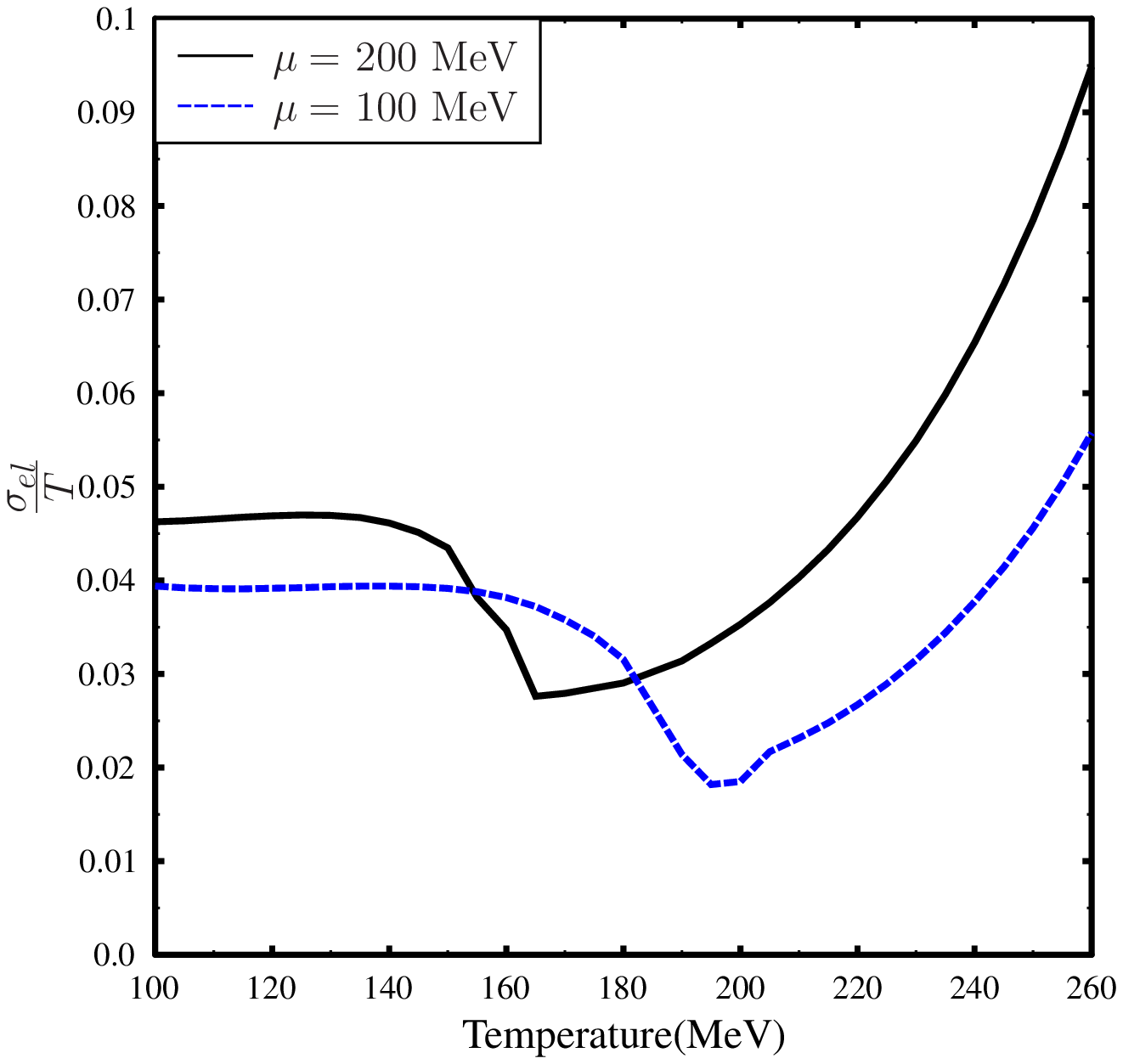}
    \end{minipage}
    \begin{minipage}{.485\textwidth}
        \centering
        \includegraphics[width=1.1\textwidth]{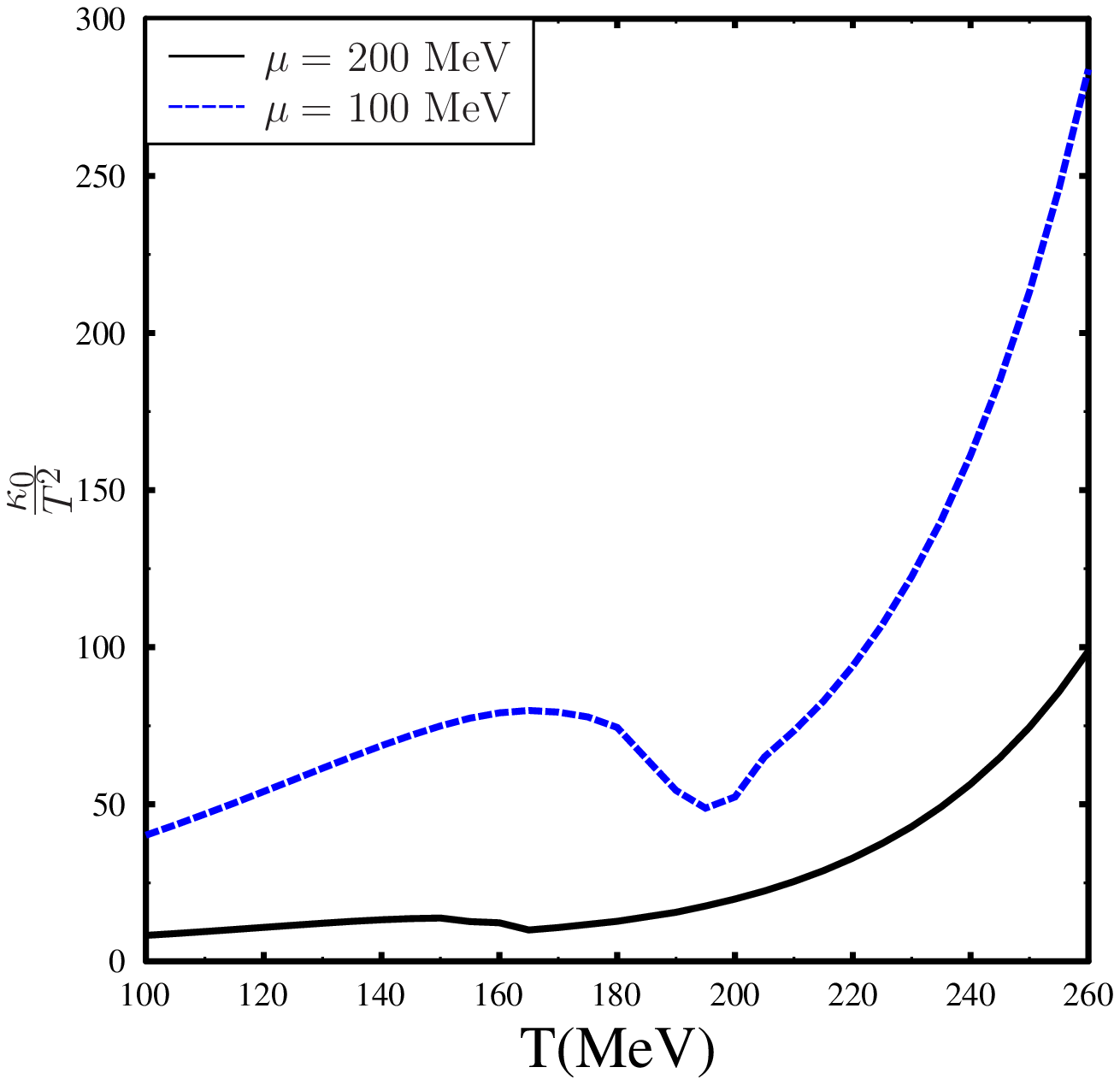}
    \end{minipage}  
    \caption{Left plot: Variation of normalized electrical conductivity ($\sigma_{el}/T$) with 
temperature for different values of the chemical potential. Right plot: Variation of normalized thermal conductivity ($\kappa_0/T^2$) with temperature for different values of the chemical potential.}
\label{fig:4}
\end{figure}

At finite quark chemical potentials, beyond the  Mott temperature, the quark-antiquark scattering still 
contributes dominantly to the scattering $\bar W_{ab}$. However, at finite densities,
there are few antiquarks as compared to quarks so that the quarks have fewer antiquarks to scatter off. 
This leads to a smaller cross-section  giving rise to a larger relaxation time for quarks compared to 
$\mu=0$ case. 
Due to the enhancement 
of quark densities at finite $\mu$, the cross-section for quark-quark scattering becomes larger resulting 
in a smaller relaxation time for the quarks compared to the case at vanishing chemical potential 
below the Mott temperature. The antiquark relaxation time, on the other hand, is always smaller 
compared to $\mu=0$ case as there are more quarks to
scatter off at finite chemical potential.

In the left plot of Fig.\eqref{fig:4} we show the behavior of normalized electrical conductivity $\sigma_{el}/T$ with temperature for different values of chemical potential. The generic behavior of relaxation time of Fig.\eqref{fig:3} is
reflected in the behavior of electrical conductivity, having a minimum at Mott transition temperature. Apart from this, it is also observed that $\sigma_{el}/T$  increases with chemical potential. This is because the contribution to the electrical
conductivity arises dominantly from quarks rather than antiquarks at finite chemical potential, as the antiquark contribution gets
suppressed due to the distribution function. This apart, there is an enhancement of the 
relaxation time at finite $\mu$ beyond the Mott transition. 
%The dominant contribution to the scattering 
%process is $u\bar d\rightarrow u\bar d$ and the $\bar d $ density
%gets suppressed as $\mu$ increase and leads to an enhancement of $\tau$. 
Both, due to an increase of dominant charge carrier densitiy and an increase in relaxation time 
with $\mu$ lead to 
enhancement of electrical conductivity beyond the Mott temperature. On the other hand, below the Mott temperature,
 although the relaxation time decrease with chemical potential for a given temperature,
 the increase in the quark number density makes 
the coefficient of electrical conductivity increasing with chemical potential.
 Further, in the high-temperature range
i.e. for temperatures much greater than the constituent quark mass $M$,
$\sigma_{el}/T$  as given in Eq.(\ref{econd}), can be shown to be $\sigma_{el}/T\sim T\tau \exp(\mu/T)$.
Therefore for a temperatures larger than the $T_M$,
$\sigma_{el}/T$ increase with temperature essentially due to increase in relaxation time. Further, at high temperatures
it increases with chemical potential due to the factor of 
 $\exp(\mu/T)$ as seen in Fig.\eqref{fig:4}.

In the right plot of Fig.\eqref{fig:4} we show the variation of the normalized thermal 
conductivity ($\kappa_0/T^2$)
with temperature. The ratio shows again a nonmonotonic variation with temperature. The origin of such behavior 
again lies with the variation of relaxation time with temperature. Beyond the Mott temperature, 
the thermal conductivity increases sharply with temperature. This can be understood as follows. For large temperatures,
when quark masses can be neglected, it can be easily shown that the enthalpy to the net quark 
number density ratio behaves as $\omega/n_q\sim T\coth(\mu/T)$. Further,  in the expression of 
the thermal conductivity as in Eq.(\ref{tcond}), $(E-\frac{\omega}{n_q})^2\sim (\frac{\omega}{n_q})^2$, due to the fact
 that single-particle energy ($E$) is much smaller than enthalpy per particle i.e. $\omega/n_q$. 
Therefore, the variation 
of the normalized thermal conductivity with temperature and chemical potential is essentially determined by the
 variation of relaxation time, $\omega/n_q$, and the distribution function with temperature and/or chemical potential. 
It can be shown as earlier, in the high-temperature limit the normalized thermal conductivity, $\kappa_0/T^2$ can
 be approximately expressed as, $\kappa_0/T^2\sim T\tau \exp(\mu/T) (\coth(\mu/T))^2$. Thus,beyond $T_M$,
 the increasing behavior of $\tau$  determines the increasing behavior of $\kappa_0/T^2$. On the other hand 
for $\mu<<T$, $\coth(\mu/T)\sim T/\mu$ in the leading order. Therefore in the high-temperature limit, 
$\kappa_0/T^2$ decreases with chemical potential. 

%the enthalpy to net quark density is proportional to $T^2/\mu$. At large temperature, in the
% expression for thermal conductivity as in Eq.(\ref{tcond}), the enthaply to quark density become much larger compared to the 
%single particle energy and on demensional ground one can show that $\kappa/T^2 \sim T^3\tau/\mu^2$. The increasing nature of $\tau$ with temperature further enhances this increasing behaviour of the ratio of $\kappa/T^2$ with temperature. Further, this also explains the decreasing behaviour of this ratio with chemical potential.

\begin{figure}
    \begin{minipage}{.485\textwidth}
        \centering
        \includegraphics[width=1.1\textwidth]{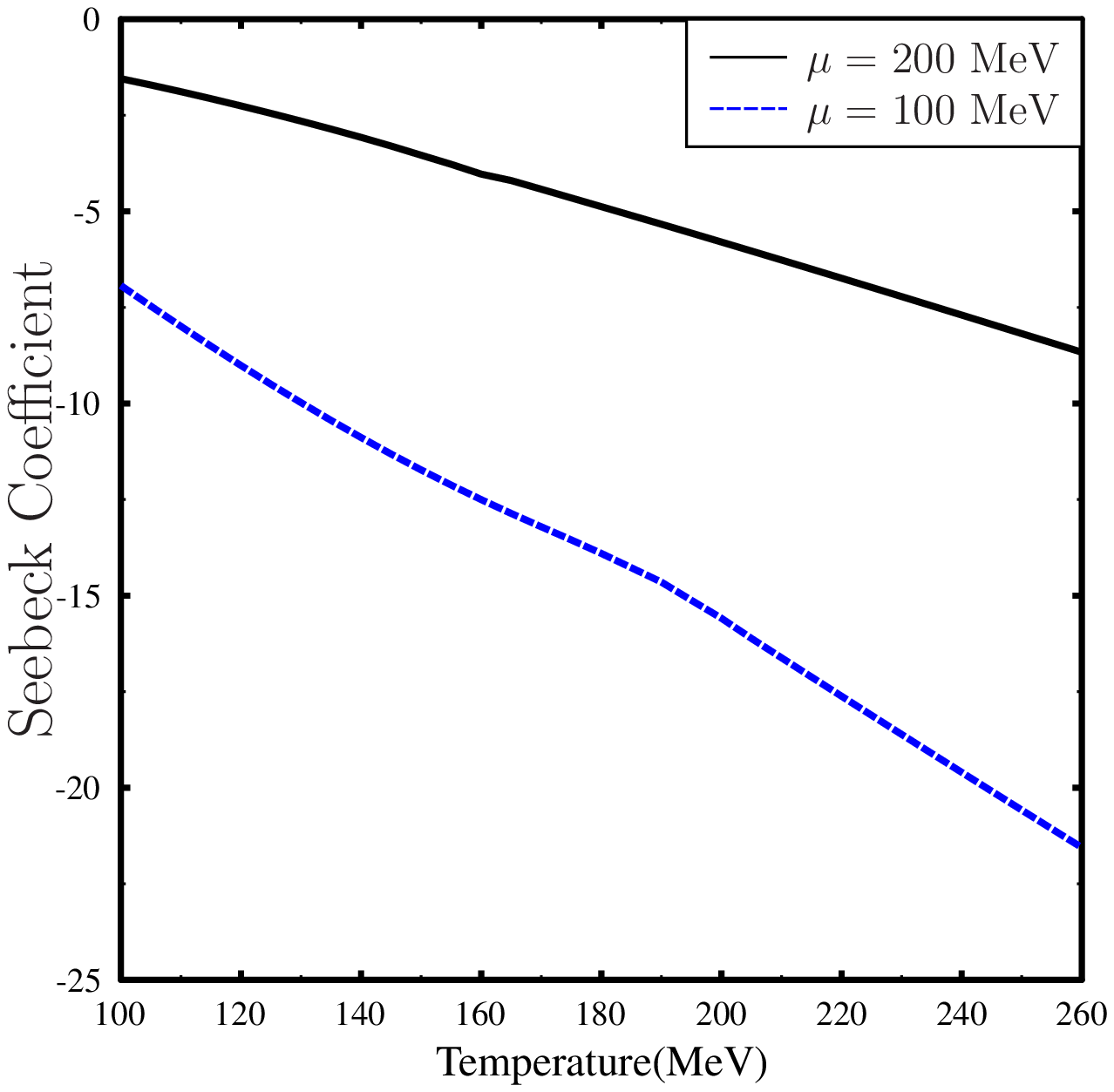}
    \end{minipage}
    \begin{minipage}{.485\textwidth}
        \centering
        \includegraphics[width=1.1\textwidth]{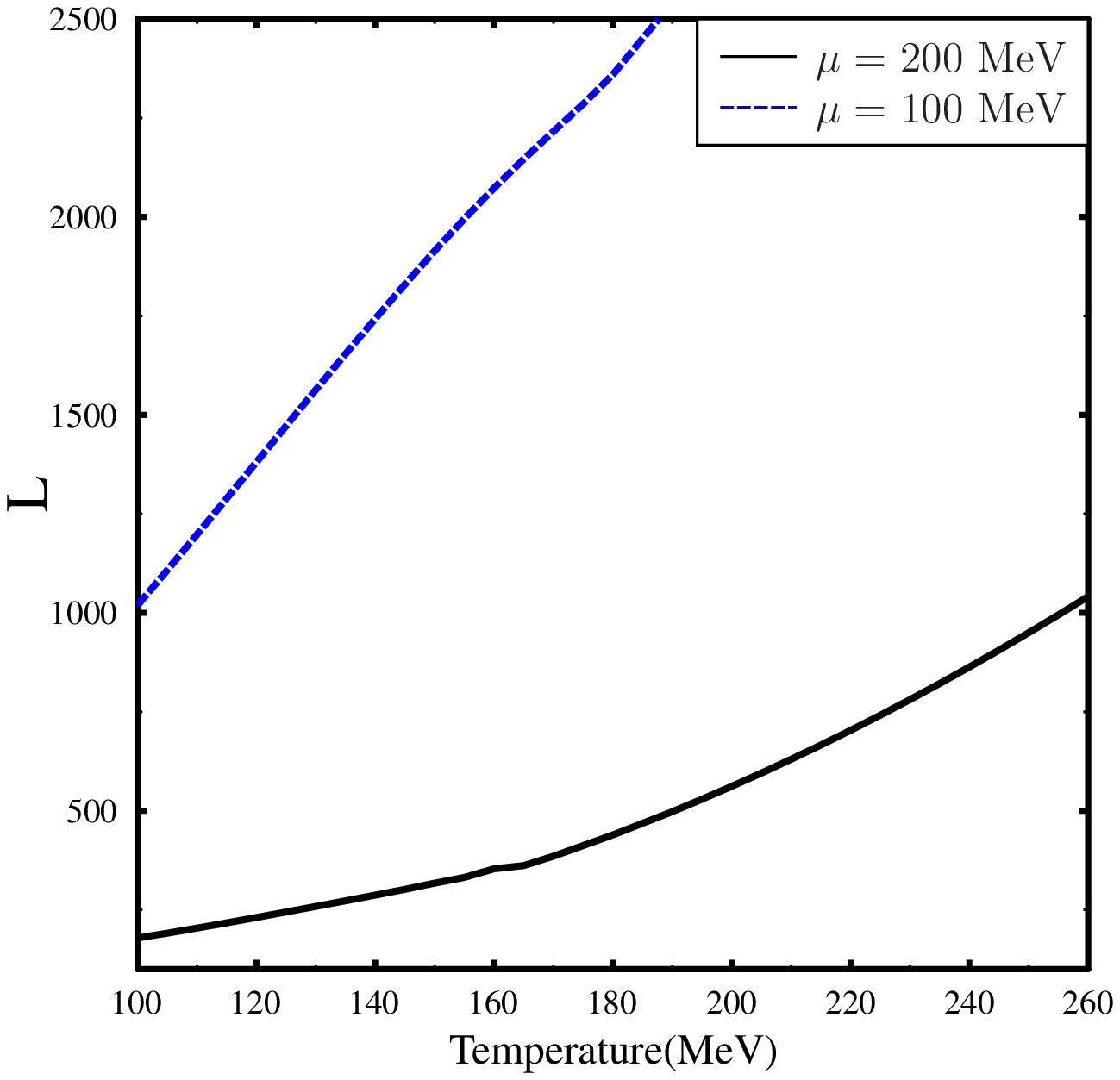}
    \end{minipage}  
    \caption{Left plot: Variation of the Seebeck coefficient with temperature for different values of the chemical potential. Right plot: Variation of the Lorenz number, $L=\kappa_0/(\sigma_{el}T)$ with temperature for different values of the chemical potential.}
 \label{fig:5}
\end{figure}

We next show the behavior of the Seebeck coefficient as a function of temperature for different values of quark chemical potential in the left plot of Fig.~\eqref{fig:5}. This coefficient, which is dimensionless, decreases monotonically with temperature. The variation of the Seebeck coefficient with temperature can be understood as follows.
First, it may be noted  that this coefficient is a ratio of two quantities each of which is proportional to
 the relaxation time.  When we consider the relaxation time as the average relaxation time, the ratio becomes 
independent of the average relaxation time. Further,  at finite chemical potential quark contribution to the 
Seebeck coefficient is dominant compared to the antiquark contribution.  
Therefore, contrary to the nonmonotonic variation of $\sigma_{el}/T$ and $\kappa_0/T^2$ with temperature, 
where the nonmonotonic variation 
 has its origin stemming from the behavior of relaxation time with temperature, the variation of the Seebeck coefficient is 
not expected to be nonmonotonic. Further, unlike other transport coefficients, the positivity of the Seebeck coefficient 
is not guaranteed. This is because in the expression of the Seebeck coefficient as given in Eq.(\ref{seeq}), 
the integrand in the numerator has the factor which is linear in $(E_a-b_a \omega/n_q)$. Therefore for the quarks, 
this factor becomes $(E-\omega/n_q)$, and the single-particle energy $E$ is much smaller than $\omega/n_q$. 
Therefore, the term $(E-\omega/n_q)$ is negative which makes the Seebeck coefficient negative. However,
 it is important to note that the expression of thermal conductivity also contains a term $(E-\omega/n_q)$,
 but it comes as a square. Therefore,the coefficient of thermal conductivity is positive define.
In  condensed matter system, the Seebeck coefficient can be both positive and negative, e.g. for electron 
and holes the Seebeck coefficients are of opposite sign.
Further, for a bipolar medium with multiple charge carriers the sign of the Seebeck coefficient depends
 on the range of temperature considered \cite{Zhou:2020}. Similar to the case of thermal conductivity, 
one can do an analysis regarding the behavior of the Seebeck coefficient in the massless limit. In the massless limit, 
it can be shown that $S\sim -\coth(\mu/T)$. Therefore for high temperatures, the leading order contribution to 
the Seebeck coefficient is $S\sim -T/\mu$. Hence with increasing temperature the Seebeck coefficient decreases, 
on the other hand with an increase in chemical potential Seebeck coefficient increases. 
% In the simple analysis,  we have assumed that the dominant contributions in the sum over species arise from quarks as the antiquark contributions are suppressed due to finite chemical potential. 

 Finally, in the right plot of Fig.~\eqref{fig:5} we have plotted the ratio $L=\kappa_0/(\sigma_{el}T)$, as a 
function of temperature. This is nothing but the Wiedemann-Franz law. In condensed matter systems,
 this ratio is a constant and is known as the Lorenz number. In the present case, however, it is observed that the 
ratio increases monotonically with temperature. Similar to the Seebeck coefficient, 
in the constant relaxation time approximation, 
the ratio $L$, is independent of relaxation time. Further, in the high temperature limit 
$\kappa_0/(\sigma T)\sim (\coth(\mu/T))^2$. Therefore in the leading order for $\mu/T$, $\kappa_0/(\sigma T)\sim T^2/\mu^2$. Hence at high temperatures the ratio $L$ increases with temperature but decreases with quark chemical potential.  
 
\section{Conclusion}
\label{conclusion}
In the present investigation, we have estimated the Seebeck coefficient in a hot and dense partonic medium modeled by the 
Nambu-Jona-Lasinio model. Here, we have considered thermoelectric effect arising from a temperature gradient as 
well as a gradient in the chemical potential. 
Apart from the Seebeck coefficient, we have also estimated electrical conductivity, thermal conductivity, and 
Lorenz number associated with the Wiedemann–Franz law. Although electrical conductivity and thermal conductivity always 
remain positive, the Seebeck coefficient is negative for the range of temperature and chemical potential considered 
here. Also the variation of electrical conductivity and thermal conductivity with temperature and quark chemical potential is intimately related to the variation of the relaxation time with temperature and chemical potential. But the variation of the Seebeck coefficient and the Lorenz number are not sensitive to the variation of relaxation time with temperature and quark chemical potential.    

In the presence of thermoelectric effects in a conducting medium, the temperature gradient can be converted into 
an electrical current and vice versa. Seebeck coefficient represents the efficiency of any conducting medium to 
convert a temperature gradient into an electrical current. Therefore, for a nonvanishing Seebeck coefficient,
the electrical current as well as the heat current get modified. The electrical current in the presence of Seebeck effect becomes, $\vec{J}=\sigma_{el} \vec{\mathcal{E}}-\sigma_{e l} S \vec{\nabla} T$.
It is important to note that the electrical conductivity $\sigma_{el}$ is always positive due to 
the contributions of both the particles and the antiparticles. Positivity of the electrical conductivity can be shown 
using entropy production i.e. second law of thermodynamics. By demanding that in the presence of electromagnetic field 
$T \partial_{\mu} s^{\mu} \geq 0,$ where $s^{\mu}$ is the entropy current, it can be shown that the
 electrical conductivity is positive definite \cite{PhysRevD.81.045015}. 
For a negative Seebeck coefficient in the presence of a positive temperature gradient the electric current 
gets enhanced. Therefore, the net electric current increases if the electric current due to the 
thermoelectric effect and the electric current due to the external electric field contributes constructively.The 
thermal conductivity in the presence of the thermoelectric effect also gets modified. In the presence of a nonvanishing
 Seebeck coefficient, the net thermal conductivity is given as $\kappa=\kappa_{0}-T \sigma_{el} S^{2}$, indicating
  the nonvanishing value of the Seebeck coefficient reduces the effective thermal conductivity. It is 
important to note that the thermal conductivity is required to be positive for the theory to be consistent with the second law of
thermodynamics, i.e., $T \partial_{\mu}s^{\mu} \geq 0$. 
Using the formalism of viscous hydrodynamics and viscous magnetohydrodynamics positivity of the electrical conductivity and the thermal conductivity has been shown explicitly \cite{PhysRevD.81.045015,Gavin:1985ph}. But the contributions to the entropy current coming from the thermoelectric effects are not considered in these investigations. Therefore in the context of entropy production in the viscous hydrodynamics and magnetohydrodynamics, it will be interesting to study the effects of thermoelectric coefficients.

Thermoelectric coefficients could also be relevant in the context of the spin Hall effect (SHE). Spin Hall effect 
is an important ingredient for the generation of spin current and it is a key concept in spintronics.
 In the generation of spin current, spin Hall effect
plays an important role. In spin Hall effect an electric field induces a transverse spin current perpendicular to the direction of the electric field. Spin Hall effect has been investigated recently in a hot and dense nuclear matter in the context of heavy-ion collisions \cite{Liu:2020dxg}. It has been argued that due to SHE, a spin current will be produced proportional to the electric field. This also means external electric field $\vec{\mathcal{E}}$ will induce a local spin polarization and the spin polarization distribution function of fermions (antifermions) in
momentum space will feature a dipole distribution. Therefore, there will be a spin flow in the
plane transverse to the direction of the electric field. Observation of
spin Hall effect may open a new direction in the exploration of the many body quantum effects in hot and dense
 nuclear matter. However, the life-time of the electric field originating in heavy-ion collisions may be 
of a small value of order 1 fm. Therefore, the idea of the observation of the spin Hall effect becomes speculative. 
However, due to the presence of nonvanishing thermoelectric coefficients, any temperature gradient as well as a gradient in the chemical potential can give rise to an effective electric field which may contribute to the spin Hall effect. Therefore a detailed analysis of the thermoelectric property of the hot and dense matter produced in a heavy ion collision experiment could be relevant for spin Hall effect and needs further investigation.

\section*{Acknowledgments}
We thank Prof. Ajit M. Srivastava for suggesting and initiating discussions on the idea of thermoelectric coefficient in the context of
heavy-ion collision. The
authors would like to thank Prof. Jitesh R. Bhatt for useful
discussions. The authors would also like to thank
Sabyasachi Ghosh, Abhishek Atreya,
Chowdhury Aminul Islam, Rajarshi Ray for many discussions on the topic of Seebeck coefficient during working
group activities at WHEPP 2017, IISER Bhopal.
The work of A.D. is supported
by the Polish National Science Center Grant No. 2018/30/E/ST2/00432.
\bibliographystyle{utphys}
\bibliography{refs.bib}

\providecommand{\href}[2]{#2}\begingroup\raggedright\begin{thebibliography}{10}

\bibitem{Heinz:2013th}
U.~Heinz and R.~Snellings, ``{Collective flow and viscosity in relativistic
  heavy-ion collisions},''
  \href{http://dx.doi.org/10.1146/annurev-nucl-102212-170540}{{\em Ann.\ Rev.\
  Nucl.\ Part.\ Sci.} {\bf 63} (2013)  123--151},
  \href{http://arxiv.org/abs/1301.2826}{{\tt arXiv:1301.2826 [nucl-th]}}.

\bibitem{Romatschke:2007mq}
P.~Romatschke and U.~Romatschke, ``{Viscosity Information from Relativistic
  Nuclear Collisions: How Perfect is the Fluid Observed at RHIC?},''
  \href{http://dx.doi.org/10.1103/PhysRevLett.99.172301}{{\em Phys.\ Rev.\
  Lett.} {\bf 99} (2007)  172301}.

\bibitem{Kovtun:2004de}
P.~Kovtun, D.~T. Son, and A.~O. Starinets, ``{Viscosity in strongly interacting
  quantum field theories from black hole physics},''
  \href{http://dx.doi.org/10.1103/PhysRevLett.94.111601}{{\em Phys.\ Rev.\
  Lett.} {\bf 94} (2005)  111601}.

\bibitem{Dobado:2012zf}
A.~Dobado and J.~M. Torres-Rincon, ``{Bulk viscosity and the phase transition
  of the linear sigma model},''
  \href{http://dx.doi.org/10.1103/PhysRevD.86.074021}{{\em Phys.\ Rev.\ D} {\bf
  86} (2012)  074021}, \href{http://arxiv.org/abs/1206.1261}{{\tt
  arXiv:1206.1261 [hep-ph]}}.

\bibitem{Sasaki:2008fg}
C.~Sasaki and K.~Redlich, ``{Bulk viscosity in quasi particle models},''
  \href{http://dx.doi.org/10.1103/PhysRevC.79.055207}{{\em Phys.\ Rev.\ C} {\bf
  79} (2009)  055207}, \href{http://arxiv.org/abs/0806.4745}{{\tt
  arXiv:0806.4745 [hep-ph]}}.

\bibitem{Sasaki:2008um}
C.~Sasaki and K.~Redlich, ``{Transport coefficients near chiral phase
  transition},'' \href{http://dx.doi.org/10.1016/j.nuclphysa.2009.11.005}{{\em
  Nucl. Phys.} {\bf A832} (2010)  62--75},
\href{http://arxiv.org/abs/0811.4708}{{\tt arXiv:0811.4708 [hep-ph]}}.
%%CITATION = ARXIV:0811.4708;%%.

\bibitem{Karsch:2007jc}
F.~Karsch, D.~Kharzeev, and K.~Tuchin, ``{Universal properties of bulk
  viscosity near the QCD phase transition},''
  \href{http://dx.doi.org/10.1016/j.physletb.2008.01.080}{{\em Phys.\ Lett.\ B}
  {\bf 663} (2008)  217--221}, \href{http://arxiv.org/abs/0711.0914}{{\tt
  arXiv:0711.0914 [hep-ph]}}.

\bibitem{Finazzo:2014cna}
S.~I. Finazzo, R.~Rougemont, H.~Marrochio, and J.~Noronha, ``{Hydrodynamic
  transport coefficients for the non-conformal quark-gluon plasma from
  holography},'' \href{http://dx.doi.org/10.1007/JHEP02(2015)051}{{\em JHEP}
  {\bf 02} (2015)  051}, \href{http://arxiv.org/abs/1412.2968}{{\tt
  arXiv:1412.2968 [hep-ph]}}.

\bibitem{Jeon:1995zm}
S.~Jeon and L.~G. Yaffe, ``{From quantum field theory to hydrodynamics:
  Transport coefficients and effective kinetic theory},''
  \href{http://dx.doi.org/10.1103/PhysRevD.53.5799}{{\em Phys. Rev. D} {\bf 53}
  (1996)  5799--5809}, \href{http://arxiv.org/abs/hep-ph/9512263}{{\tt
  arXiv:hep-ph/9512263}}.

\bibitem{Bazavov:2009zn}
A.~Bazavov {\em et al.}, ``Equation of state and qcd transition at finite
  temperature,'' \href{http://dx.doi.org/10.1103/PhysRevD.80.014504}{{\em Phys.
  Rev. D} {\bf 80} (2009)  014504}, \href{http://arxiv.org/abs/0903.4379}{{\tt
  arXiv:0903.4379 [hep-lat]}}.

\bibitem{Bazavov:2010pg}
{\bf HotQCD} Collaboration, A.~Bazavov and P.~Petreczky, ``Taste symmetry and
  qcd thermodynamics with improved staggered fermions,''
  \href{http://dx.doi.org/10.22323/1.105.0169}{{\em PoS} {\bf LATTICE2010}
  (2010)  169}, \href{http://arxiv.org/abs/1012.1257}{{\tt arXiv:1012.1257
  [hep-lat]}}.

\bibitem{Bozek:2009dw}
P.~Bozek, ``{Bulk and shear viscosities of matter created in relativistic
  heavy-ion collisions},''
  \href{http://dx.doi.org/10.1103/PhysRevC.81.034909}{{\em Phys. Rev. C} {\bf
  81} (2010)  034909}, \href{http://arxiv.org/abs/0911.2397}{{\tt
  arXiv:0911.2397 [nucl-th]}}.

\bibitem{Rose:2014fba}
J.-B. Rose, J.-F. Paquet, G.~S. Denicol, M.~Luzum, B.~Schenke, S.~Jeon, and
  C.~Gale, ``{Extracting the bulk viscosity of the quark--gluon plasma},''
  \href{http://dx.doi.org/10.1016/j.nuclphysa.2014.09.044}{{\em Nucl. Phys. A}
  {\bf 931} (2014)  926--930}, \href{http://arxiv.org/abs/1408.0024}{{\tt
  arXiv:1408.0024 [nucl-th]}}.

\bibitem{Tuchin:2010gx}
K.~Tuchin, ``{Photon decay in strong magnetic field in heavy-ion collisions},''
  \href{http://dx.doi.org/10.1103/PhysRevC.83.017901}{{\em Phys.\ Rev.\ C} {\bf
  83} (2011)  017901}, \href{http://arxiv.org/abs/1008.1604}{{\tt
  arXiv:1008.1604 [nucl-th]}}.

\bibitem{Tuchin:2010vs}
K.~Tuchin, ``{Synchrotron radiation by fast fermions in heavy-ion
  collisions},'' \href{http://dx.doi.org/10.1103/PhysRevC.83.039903,
  10.1103/PhysRevC.82.034904}{{\em Phys. Rev.} {\bf C82} (2010)  034904},
  \href{http://arxiv.org/abs/1006.3051}{{\tt arXiv:1006.3051 [nucl-th]}}.
[Erratum: Phys. Rev.C83,039903(2011)].
%%CITATION = ARXIV:1006.3051;%%.

\bibitem{Inghirami:2016iru}
G.~Inghirami, L.~Del~Zanna, A.~Beraudo, M.~H. Moghaddam, F.~Becattini, and
  M.~Bleicher, ``{Numerical magneto-hydrodynamics for relativistic nuclear
  collisions},'' \href{http://dx.doi.org/10.1140/epjc/s10052-016-4516-8}{{\em
  Eur. Phys. J. C} {\bf 76} (2016) no.~12, 659},
  \href{http://arxiv.org/abs/1609.03042}{{\tt arXiv:1609.03042 [hep-ph]}}.

\bibitem{Das:2017qfi}
A.~Das, S.~S. Dave, P.~Saumia, and A.~M. Srivastava, ``{Effects of magnetic
  field on plasma evolution in relativistic heavy-ion collisions},''
  \href{http://dx.doi.org/10.1103/PhysRevC.96.034902}{{\em Phys.\ Rev.\ C} {\bf
  96} (2017) no.~3, 034902}, \href{http://arxiv.org/abs/1703.08162}{{\tt
  arXiv:1703.08162 [hep-ph]}}.

\bibitem{Greif:2016skc}
M.~Greif, C.~Greiner, and G.~S. Denicol, ``{Electric conductivity of a hot
  hadron gas from a kinetic approach},''
  \href{http://dx.doi.org/10.1103/PhysRevD.93.096012}{{\em Phys. Rev. D} {\bf
  93} (2016) no.~9, 096012}, \href{http://arxiv.org/abs/1602.05085}{{\tt
  arXiv:1602.05085 [nucl-th]}}. [Erratum: Phys.Rev.D 96, 059902 (2017)].

\bibitem{Greif:2014oia}
M.~Greif, I.~Bouras, C.~Greiner, and Z.~Xu, ``{Electric conductivity of the
  quark-gluon plasma investigated using a perturbative QCD based parton
  cascade},'' \href{http://dx.doi.org/10.1103/PhysRevD.90.094014}{{\em Phys.\
  Rev.\ D} {\bf 90} (2014) no.~9, 094014},
  \href{http://arxiv.org/abs/1408.7049}{{\tt arXiv:1408.7049 [nucl-th]}}.

\bibitem{Puglisi:2014pda}
A.~Puglisi, S.~Plumari, and V.~Greco, ``{Shear viscosity $\eta$ to electric
  conductivity $\sigma_{el}$ ratio for the quark–gluon plasma},''
  \href{http://dx.doi.org/10.1016/j.physletb.2015.10.070}{{\em Phys. Lett.}
  {\bf B751} (2015)  326--330},
\href{http://arxiv.org/abs/1407.2559}{{\tt arXiv:1407.2559 [hep-ph]}}.
%%CITATION = ARXIV:1407.2559;%%.

\bibitem{Puglisi:2014sha}
A.~Puglisi, S.~Plumari, and V.~Greco, ``{Electric Conductivity from the
  solution of the Relativistic Boltzmann Equation},''
  \href{http://dx.doi.org/10.1103/PhysRevD.90.114009}{{\em Phys.\ Rev.\ D} {\bf
  90} (2014)  114009}, \href{http://arxiv.org/abs/1408.7043}{{\tt
  arXiv:1408.7043 [hep-ph]}}.

\bibitem{Cassing:2013iz}
W.~Cassing, O.~Linnyk, T.~Steinert, and V.~Ozvenchuk, ``{Electrical
  Conductivity of Hot QCD Matter},''
  \href{http://dx.doi.org/10.1103/PhysRevLett.110.182301}{{\em Phys.\ Rev.\
  Lett.} {\bf 110} (2013) no.~18, 182301},
  \href{http://arxiv.org/abs/1302.0906}{{\tt arXiv:1302.0906 [hep-ph]}}.

\bibitem{Steinert:2013fza}
T.~Steinert and W.~Cassing, ``{Electric and magnetic response of hot QCD
  matter},'' \href{http://dx.doi.org/10.1103/PhysRevC.89.035203}{{\em Phys.\
  Rev.\ C} {\bf 89} (2014) no.~3, 035203},
  \href{http://arxiv.org/abs/1312.3189}{{\tt arXiv:1312.3189 [hep-ph]}}.

\bibitem{Aarts:2014nba}
G.~Aarts, C.~Allton, A.~Amato, P.~Giudice, S.~Hands, and J.-I. Skullerud,
  ``{Electrical conductivity and charge diffusion in thermal QCD from the
  lattice},'' \href{http://dx.doi.org/10.1007/JHEP02(2015)186}{{\em JHEP} {\bf
  02} (2015)  186}, \href{http://arxiv.org/abs/1412.6411}{{\tt arXiv:1412.6411
  [hep-lat]}}.

\bibitem{Aarts:2007wj}
G.~Aarts, C.~Allton, J.~Foley, S.~Hands, and S.~Kim, ``{Spectral functions at
  small energies and the electrical conductivity in hot, quenched lattice
  QCD},'' \href{http://dx.doi.org/10.1103/PhysRevLett.99.022002}{{\em Phys.\
  Rev.\ Lett.} {\bf 99} (2007)  022002},
  \href{http://arxiv.org/abs/hep-lat/0703008}{{\tt arXiv:hep-lat/0703008}}.

\bibitem{Amato:2013naa}
A.~Amato, G.~Aarts, C.~Allton, P.~Giudice, S.~Hands, and J.-I. Skullerud,
  ``{Electrical conductivity of the quark-gluon plasma across the deconfinement
  transition},'' \href{http://dx.doi.org/10.1103/PhysRevLett.111.172001}{{\em
  Phys.\ Rev.\ Lett.} {\bf 111} (2013) no.~17, 172001},
  \href{http://arxiv.org/abs/1307.6763}{{\tt arXiv:1307.6763 [hep-lat]}}.

\bibitem{Gupta:2003zh}
S.~Gupta, ``{The Electrical conductivity and soft photon emissivity of the QCD
  plasma},'' \href{http://dx.doi.org/10.1016/j.physletb.2004.05.079}{{\em
  Phys.\ Lett.\ B} {\bf 597} (2004)  57--62},
  \href{http://arxiv.org/abs/hep-lat/0301006}{{\tt arXiv:hep-lat/0301006}}.

\bibitem{Ding:2010ga}
H.-T. Ding, A.~Francis, O.~Kaczmarek, F.~Karsch, E.~Laermann, and W.~Soeldner,
  ``{Thermal dilepton rate and electrical conductivity: An analysis of vector
  current correlation functions in quenched lattice QCD},''
  \href{http://dx.doi.org/10.1103/PhysRevD.83.034504}{{\em Phys.\ Rev.\ D} {\bf
  83} (2011)  034504}, \href{http://arxiv.org/abs/1012.4963}{{\tt
  arXiv:1012.4963 [hep-lat]}}.

\bibitem{Kaczmarek:2013dya}
O.~Kaczmarek and M.~Müller, ``{Temperature dependence of electrical
  conductivity and dilepton rates from hot quenched lattice QCD},''
  \href{http://dx.doi.org/10.22323/1.187.0175}{{\em PoS} {\bf LATTICE2013}
  (2014)  175}, \href{http://arxiv.org/abs/1312.5609}{{\tt arXiv:1312.5609
  [hep-lat]}}.

\bibitem{Qin:2013aaa}
S.-x. Qin, ``{A divergence-free method to extract observables from correlation
  functions},'' \href{http://dx.doi.org/10.1016/j.physletb.2015.02.009}{{\em
  Phys.\ Lett.\ B} {\bf 742} (2015)  358--362},
  \href{http://arxiv.org/abs/1307.4587}{{\tt arXiv:1307.4587 [nucl-th]}}.

\bibitem{Marty:2013ita}
R.~Marty, E.~Bratkovskaya, W.~Cassing, J.~Aichelin, and H.~Berrehrah,
  ``{Transport coefficients from the Nambu-Jona-Lasinio model for $SU(3)_f$},''
  \href{http://dx.doi.org/10.1103/PhysRevC.88.045204}{{\em Phys. Rev.} {\bf
  C88} (2013)  045204},
\href{http://arxiv.org/abs/1305.7180}{{\tt arXiv:1305.7180 [hep-ph]}}.
%%CITATION = ARXIV:1305.7180;%%.

\bibitem{FernandezFraile:2005ka}
D.~Fernandez-Fraile and A.~Gomez~Nicola, ``{The Electrical conductivity of a
  pion gas},'' \href{http://dx.doi.org/10.1103/PhysRevD.73.045025}{{\em Phys.\
  Rev.\ D} {\bf 73} (2006)  045025},
  \href{http://arxiv.org/abs/hep-ph/0512283}{{\tt arXiv:hep-ph/0512283}}.

\bibitem{Kharzeev:2007jp}
D.~E. Kharzeev, L.~D. McLerran, and H.~J. Warringa, ``The effects of
  topological charge change in heavy ion collisions: 'event by event p and cp
  violation','' \href{http://dx.doi.org/10.1016/j.nuclphysa.2008.02.298}{{\em
  Nucl. Phys. A} {\bf 803} (2008)  227--253},
  \href{http://arxiv.org/abs/0711.0950}{{\tt arXiv:0711.0950 [hep-ph]}}.

\bibitem{Skokov:2009qp}
V.~Skokov, A.~Illarionov, and V.~Toneev, ``Estimate of the magnetic field
  strength in heavy-ion collisions,''
  \href{http://dx.doi.org/10.1142/S0217751X09047570}{{\em Int. J. Mod. Phys. A}
  {\bf 24} (2009)  5925--5932}, \href{http://arxiv.org/abs/0907.1396}{{\tt
  arXiv:0907.1396 [nucl-th]}}.

\bibitem{Li:2016tel}
H.~Li, X.-l. Sheng, and Q.~Wang, ``Electromagnetic fields with electric and
  chiral magnetic conductivities in heavy ion collisions,''
  \href{http://dx.doi.org/10.1103/PhysRevC.94.044903}{{\em Phys. Rev. C} {\bf
  94} (2016) no.~4, 044903}, \href{http://arxiv.org/abs/1602.02223}{{\tt
  arXiv:1602.02223 [nucl-th]}}.

\bibitem{Inghirami:2019mkc}
G.~Inghirami, M.~Mace, Y.~Hirono, L.~Del~Zanna, D.~E. Kharzeev, and
  M.~Bleicher, ``Magnetic fields in heavy ion collisions: flow and charge
  transport,'' \href{http://dx.doi.org/10.1140/epjc/s10052-020-7847-4}{{\em
  Eur. Phys. J. C} {\bf 80} (2020) no.~3, 293},
  \href{http://arxiv.org/abs/1908.07605}{{\tt arXiv:1908.07605 [hep-ph]}}.

\bibitem{Inghirami:2018ziv}
G.~Inghirami, L.~Del~Zanna, A.~Beraudo, M.~Haddadi~Moghaddam, F.~Becattini, and
  M.~Bleicher, ``Magneto-hydrodynamic simulations of heavy ion collisions with
  echo-qgp,'' \href{http://dx.doi.org/10.1088/1742-6596/1024/1/012043}{{\em J.
  Phys. Conf. Ser.} {\bf 1024} (2018) no.~1, 012043}.

\bibitem{Shokri:2017xxn}
M.~Shokri and N.~Sadooghi, ``Novel self-similar rotating solutions of nonideal
  transverse magnetohydrodynamics,''
  \href{http://dx.doi.org/10.1103/PhysRevD.96.116008}{{\em Phys. Rev. D} {\bf
  96} (2017) no.~11, 116008}, \href{http://arxiv.org/abs/1705.00536}{{\tt
  arXiv:1705.00536 [nucl-th]}}.

\bibitem{Shokri:2018qcu}
M.~Shokri and N.~Sadooghi, ``Evolution of magnetic fields from the 3 + 1
  dimensional self-similar and gubser flows in ideal relativistic
  magnetohydrodynamics,'' \href{http://dx.doi.org/10.1007/JHEP11(2018)181}{{\em
  JHEP} {\bf 11} (2018)  181}, \href{http://arxiv.org/abs/1807.09487}{{\tt
  arXiv:1807.09487 [nucl-th]}}.

\bibitem{Tabatabaee:2020efb}
S.~Tabatabaee and N.~Sadooghi, ``Wigner function formalism and the evolution of
  thermodynamic quantities in an expanding magnetized plasma,''
  \href{http://dx.doi.org/10.1103/PhysRevD.101.076022}{{\em Phys. Rev. D} {\bf
  101} (2020) no.~7, 076022}, \href{http://arxiv.org/abs/2003.01686}{{\tt
  arXiv:2003.01686 [hep-ph]}}.

\bibitem{Kharzeev:2012ph}
D.~E. Kharzeev, K.~Landsteiner, A.~Schmitt, and H.-U. Yee, {\em {'Strongly
  interacting matter in magnetic fields': an overview}}, vol.~871,
  \href{http://dx.doi.org/10.1007/978-3-642-37305-3\_1}{pp.~1--11}.
\newblock 2013.
\newblock \href{http://arxiv.org/abs/1211.6245}{{\tt arXiv:1211.6245
  [hep-ph]}}.

\bibitem{Greif:2017byw}
M.~Greif, J.~A. Fotakis, G.~S. Denicol, and C.~Greiner, ``{Diffusion of
  conserved charges in relativistic heavy ion collisions},''
  \href{http://dx.doi.org/10.1103/PhysRevLett.120.242301}{{\em Phys. Rev.
  Lett.} {\bf 120} (2018) no.~24, 242301},
  \href{http://arxiv.org/abs/1711.08680}{{\tt arXiv:1711.08680 [hep-ph]}}.

\bibitem{Prakash:1993bt}
M.~Prakash, M.~Prakash, R.~Venugopalan, and G.~Welke, ``{Nonequilibrium
  properties of hadronic mixtures},''
  \href{http://dx.doi.org/10.1016/0370-1573(93)90092-R}{{\em Phys.\ Rept.} {\bf
  227} (1993)  321--366}.

\bibitem{Wiranata:2012br}
A.~Wiranata and M.~Prakash, ``{Shear Viscosities from the Chapman-Enskog and
  the Relaxation Time Approaches},''
  \href{http://dx.doi.org/10.1103/PhysRevC.85.054908}{{\em Phys.\ Rev.\ C} {\bf
  85} (2012)  054908}, \href{http://arxiv.org/abs/1203.0281}{{\tt
  arXiv:1203.0281 [nucl-th]}}.

\bibitem{Chakraborty:2010fr}
P.~Chakraborty and J.~Kapusta, ``{Quasi-Particle Theory of Shear and Bulk
  Viscosities of Hadronic Matter},''
  \href{http://dx.doi.org/10.1103/PhysRevC.83.014906}{{\em Phys.\ Rev.\ C} {\bf
  83} (2011)  014906}, \href{http://arxiv.org/abs/1006.0257}{{\tt
  arXiv:1006.0257 [nucl-th]}}.

\bibitem{Khvorostukhin:2010aj}
A.~Khvorostukhin, V.~Toneev, and D.~Voskresensky, ``{Viscosity Coefficients for
  Hadron and Quark-Gluon Phases},''
  \href{http://dx.doi.org/10.1016/j.nuclphysa.2010.05.058}{{\em Nucl. Phys. A}
  {\bf 845} (2010)  106--146}, \href{http://arxiv.org/abs/1003.3531}{{\tt
  arXiv:1003.3531 [nucl-th]}}.

\bibitem{Plumari:2012ep}
S.~Plumari, A.~Puglisi, F.~Scardina, and V.~Greco, ``{Shear Viscosity of a
  strongly interacting system: Green-Kubo vs. Chapman-Enskog and Relaxation
  Time Approximation},''
  \href{http://dx.doi.org/10.1103/PhysRevC.86.054902}{{\em Phys.\ Rev.\ C} {\bf
  86} (2012)  054902}, \href{http://arxiv.org/abs/1208.0481}{{\tt
  arXiv:1208.0481 [nucl-th]}}.

\bibitem{Gorenstein:2007mw}
M.~Gorenstein, M.~Hauer, and O.~Moroz, ``{Viscosity in the excluded volume
  hadron gas model},'' \href{http://arxiv.org/abs/0708.0137}{{\tt
  arXiv:0708.0137 [nucl-th]}}.

\bibitem{NoronhaHostler:2012ug}
J.~Noronha-Hostler, J.~Noronha, and C.~Greiner, ``{Hadron Mass Spectrum and the
  Shear Viscosity to Entropy Density Ratio of Hot Hadronic Matter},''
  \href{http://dx.doi.org/10.1103/PhysRevC.86.024913}{{\em Phys. Rev. C} {\bf
  86} (2012)  024913}, \href{http://arxiv.org/abs/1206.5138}{{\tt
  arXiv:1206.5138 [nucl-th]}}.

\bibitem{Tiwari:2011km}
S.~Tiwari, P.~Srivastava, and C.~Singh, ``{Description of Hot and Dense Hadron
  Gas Properties in a New Excluded-Volume model},''
  \href{http://dx.doi.org/10.1103/PhysRevC.85.014908}{{\em Phys.\ Rev.\ C} {\bf
  85} (2012)  014908}, \href{http://arxiv.org/abs/1111.2406}{{\tt
  arXiv:1111.2406 [hep-ph]}}.

\bibitem{Ghosh:2013cba}
S.~Ghosh, A.~Lahiri, S.~Majumder, R.~Ray, and S.~K. Ghosh, ``{Shear viscosity
  due to Landau damping from the quark-pion interaction},''
  \href{http://dx.doi.org/10.1103/PhysRevC.88.068201}{{\em Phys.\ Rev.\ C} {\bf
  88} (2013) no.~6, 068201}, \href{http://arxiv.org/abs/1311.4070}{{\tt
  arXiv:1311.4070 [nucl-th]}}.

\bibitem{Lang:2015nca}
R.~Lang, N.~Kaiser, and W.~Weise, ``{Shear viscosities from Kubo formalism in a
  large-N$_{c}$ Nambu-Jona-Lasinio model},''
  \href{http://dx.doi.org/10.1140/epja/i2015-15127-7}{{\em Eur.\ Phys.\ J.\ A}
  {\bf 51} (2015) no.~10, 127}, \href{http://arxiv.org/abs/1506.02459}{{\tt
  arXiv:1506.02459 [hep-ph]}}.

\bibitem{Ghosh:2014qba}
S.~Ghosh, G.~Krein, and S.~Sarkar, ``{Shear viscosity of a pion gas resulting
  from $\rho\pi\pi$ and $\sigma\pi\pi$ interactions},''
  \href{http://dx.doi.org/10.1103/PhysRevC.89.045201}{{\em Phys.\ Rev.\ C} {\bf
  89} (2014) no.~4, 045201}, \href{http://arxiv.org/abs/1401.5392}{{\tt
  arXiv:1401.5392 [nucl-th]}}.

\bibitem{Wiranata:2014kva}
A.~Wiranata, V.~Koch, M.~Prakash, and X.~Wang, ``{Shear viscosity of a
  multi-component hadronic system},''
  \href{http://dx.doi.org/10.1088/1742-6596/509/1/012049}{{\em J.\ Phys.\
  Conf.\ Ser.} {\bf 509} (2014)  012049}.

\bibitem{Wiranata:2012vv}
A.~Wiranata, M.~Prakash, and P.~Chakraborty, ``{Comparison of Viscosities from
  the Chapman-Enskog and Relaxation Time Methods},''
  \href{http://dx.doi.org/10.2478/s11534-012-0082-3}{{\em Central Eur. J.
  Phys.} {\bf 10} (2012)  1349--1351},
  \href{http://arxiv.org/abs/1201.3104}{{\tt arXiv:1201.3104 [nucl-th]}}.

\bibitem{NoronhaHostler:2008ju}
J.~Noronha-Hostler, J.~Noronha, and C.~Greiner, ``{Transport Coefficients of
  Hadronic Matter near T(c)},''
  \href{http://dx.doi.org/10.1103/PhysRevLett.103.172302}{{\em Phys.\ Rev.\
  Lett.} {\bf 103} (2009)  172302}, \href{http://arxiv.org/abs/0811.1571}{{\tt
  arXiv:0811.1571 [nucl-th]}}.

\bibitem{Kadam:2014cua}
G.~P. Kadam and H.~Mishra, ``{Bulk and shear viscosities of hot and dense
  hadron gas},'' \href{http://dx.doi.org/10.1016/j.nuclphysa.2014.12.004}{{\em
  Nucl. Phys. A} {\bf 934} (2014)  133--147},
  \href{http://arxiv.org/abs/1408.6329}{{\tt arXiv:1408.6329 [hep-ph]}}.

\bibitem{Kadam:2014xka}
G.~Kadam, ``{Transport properties of hadronic matter in magnetic field},''
  \href{http://dx.doi.org/10.1142/S0217732315500315}{{\em Mod.\ Phys.\ Lett.\
  A} {\bf 30} (2015) no.~10, 1550031},
  \href{http://arxiv.org/abs/1412.5303}{{\tt arXiv:1412.5303 [hep-ph]}}.

\bibitem{Ghosh:2014yea}
S.~Ghosh, ``{A real-time thermal field theoretical analysis of Kubo-type shear
  viscosity: Numerical understanding with simple examples},''
  \href{http://dx.doi.org/10.1142/S0217751X14500547}{{\em Int.\ J.\ Mod.\
  Phys.\ A} {\bf 29} (2014)  1450054},
  \href{http://arxiv.org/abs/1404.4788}{{\tt arXiv:1404.4788 [nucl-th]}}.

\bibitem{Rose:2017bjz}
J.~B. Rose, J.~Torres-Rincon, A.~Schäfer, D.~Oliinychenko, and H.~Petersen,
  ``{Shear viscosity of a hadron gas and influence of resonance lifetimes on
  relaxation time},'' \href{http://dx.doi.org/10.1103/PhysRevC.97.055204}{{\em
  Phys.\ Rev.\ C} {\bf 97} (2018) no.~5, 055204},
  \href{http://arxiv.org/abs/1709.03826}{{\tt arXiv:1709.03826 [nucl-th]}}.

\bibitem{Wesp:2011yy}
C.~Wesp, A.~El, F.~Reining, Z.~Xu, I.~Bouras, and C.~Greiner, ``{Calculation of
  shear viscosity using Green-Kubo relations within a parton cascade},''
  \href{http://dx.doi.org/10.1103/PhysRevC.84.054911}{{\em Phys.\ Rev.\ C} {\bf
  84} (2011)  054911}, \href{http://arxiv.org/abs/1106.4306}{{\tt
  arXiv:1106.4306 [hep-ph]}}.

\bibitem{Denicol:2012vq}
G.~Denicol, H.~Niemi, I.~Bouras, E.~Molnar, Z.~Xu, D.~Rischke, and C.~Greiner,
  ``{Solving the heat-flow problem with transient relativistic fluid
  dynamics},'' \href{http://dx.doi.org/10.1103/PhysRevD.89.074005}{{\em Phys.\
  Rev.\ D} {\bf 89} (2014) no.~7, 074005},
  \href{http://arxiv.org/abs/1207.6811}{{\tt arXiv:1207.6811 [nucl-th]}}.

\bibitem{Kapusta:2012zb}
J.~I. Kapusta and J.~M. Torres-Rincon, ``{Thermal Conductivity and Chiral
  Critical Point in Heavy Ion Collisions},''
  \href{http://dx.doi.org/10.1103/PhysRevC.86.054911}{{\em Phys.\ Rev.\ C} {\bf
  86} (2012)  054911}, \href{http://arxiv.org/abs/1209.0675}{{\tt
  arXiv:1209.0675 [nucl-th]}}.

\bibitem{seebconds1}
P.~Ao, ``{Nernst Effect, Seebeck Effect, and Vortex Dynamics in the Mixed State
  of Superconductors},'' {\em arXiv: cond-mat/9505002}  .

\bibitem{seebconds2}
M.~Matusiak, K.~Rogacki, and T.~Wolf, ``{Thermoelectric anisotropy in the
  iron-based superconductor
  $\mathrm{Ba}{({\mathrm{Fe}}_{1\ensuremath{-}x}{\mathrm{Co}}_{x})}_{2}{\mathrm{As}}_{2}$},''
  {\em Phys. Rev. B} {\bf 97} (2018)  220501.

\bibitem{seebconds3}
C.~S.~Y. M.~K.~Hooda, ``Electronic transport properties of intermediately
  coupled superconductors: Pdte2 and cu0.04pdte2,'' {\em arXiv:1704.07194}  .

\bibitem{seebconds4}
O.~Cyr-Choini\`ere, S.~Badoux, G.~Grissonnanche, B.~Michon, S.~A.~A. Afshar,
  S.~Fortier, D.~LeBoeuf, D.~Graf, J.~Day, D.~A. Bonn, W.~N. Hardy, R.~Liang,
  N.~Doiron-Leyraud, and L.~Taillefer, ``Anisotropy of the seebeck coefficient
  in the cuprate superconductor
  ${\mathrm{yba}}_{2}{\mathrm{cu}}_{3}{\mathrm{o}}_{y}$: Fermi-surface
  reconstruction by bidirectional charge order,'' {\em Phys. Rev. X} {\bf 7}
  (2017)  031042.

\bibitem{seebconds5}
S.~Sergei, ``Nonlinear seebeck effect in a model granular superconductor,''
  {\em JETP Letters} {\bf 67} (1998)  680--684.

\bibitem{seebconds6}
M.~M. Wysokinski and J.~Spalek, ``Seebeck effect in the graphene-superconductor
  junction,'' {\em Journal of Applied Physics} {\bf 113} (2013)  163905.

\bibitem{seebconds7}
K.~P. W\'ojcik and I.~Weymann, ``Proximity effect on spin-dependent conductance
  and thermopower of correlated quantum dots,'' {\em Phys. Rev. B} {\bf 89}
  (2014)  165303.

\bibitem{seebconds8}
K.~Seo and S.~Tewari, ``Fermi-surface reconstruction and transport coefficients
  from a mean-field bidirectional charge-density wave state in the
  high-${T}_{c}$ cuprates,'' {\em Phys. Rev. B} {\bf 90} (2014)  174503.

\bibitem{seebconds9}
P.~Dutta, A.~Saha, and A.~M. Jayannavar, ``Thermoelectric properties of a
  ferromagnet-superconductor hybrid junction: Role of interfacial rashba
  spin-orbit interaction,'' {\em Phys. Rev. B} {\bf 96} (2017)  115404.

\bibitem{Das:2020beh}
A.~Das, H.~Mishra, and R.~K. Mohapatra, ``{Magneto-Seebeck coefficient and
  Nernst coefficient of hot and dense hadron gas},'' {\em arXiv: 2004.04665}  .

\bibitem{Bhatt:2018ncr}
J.~R. Bhatt, A.~Das, and H.~Mishra, ``{Thermoelectric effect and Seebeck
  coefficient for hot and dense hadronic matter},''
  \href{http://dx.doi.org/10.1103/PhysRevD.99.014015}{{\em Phys. Rev. D} {\bf
  99} (2019) no.~1, 014015}.

\bibitem{Dey:2020sbm}
D.~Dey and B.~K. Patra, ``{Seebeck effect in a thermal QCD medium in the
  presence of strong magnetic field},'' {\em arXiv: 2004.03149}  .

\bibitem{Zhang:2020efz}
H.-X. Zhang, ``{Thermoelectric properties of (an-)isotropic QGP in magnetic
  fields},'' {\em arXiv:2004.08767}  .

\bibitem{Singha:2017jmq}
P.~Singha, A.~Abhishek, G.~Kadam, S.~Ghosh, and H.~Mishra, ``{Calculations of
  shear, bulk viscosities and electrical conductivity in the
  Polyakov-quark--meson model},''
  \href{http://dx.doi.org/10.1088/1361-6471/aaf256}{{\em J. Phys. G} {\bf 46}
  (2019) no.~1, 015201}.

\bibitem{Abhishek:2017pkp}
A.~Abhishek, H.~Mishra, and S.~Ghosh, ``{Transport coefficients in the Polyakov
  quark meson coupling model: A relaxation time approximation},''
  \href{http://dx.doi.org/10.1103/PhysRevD.97.014005}{{\em Phys. Rev. D} {\bf
  97} (2018) no.~1, 014005}.

\bibitem{Singh:2018wps}
B.~Singh, A.~Abhishek, S.~K. Das, and H.~Mishra, ``{Heavy quark diffusion in a
  Polyakov loop plasma},''
  \href{http://dx.doi.org/10.1103/PhysRevD.100.114019}{{\em Phys. Rev. D} {\bf
  100} (2019) no.~11, 114019}.

\bibitem{Ratti:2005jh}
C.~Ratti, M.~A. Thaler, and W.~Weise, ``{Phases of QCD: Lattice thermodynamics
  and a field theoretical model},''
  \href{http://dx.doi.org/10.1103/PhysRevD.73.014019}{{\em Phys. Rev. D} {\bf
  73} (2006)  014019}.

\bibitem{Deb:2016myz}
P.~Deb, G.~P. Kadam, and H.~Mishra, ``{Estimating transport coefficients in hot
  and dense quark matter},''
  \href{http://dx.doi.org/10.1103/PhysRevD.94.094002}{{\em Phys. Rev.} {\bf
  D94} (2016) no.~9, 094002},
\href{http://arxiv.org/abs/1603.01952}{{\tt arXiv:1603.01952 [hep-ph]}}.
%%CITATION = ARXIV:1603.01952;%%.

\bibitem{Rehberg:1995kh}
P.~Rehberg, S.~P. Klevansky, and J.~Hufner, ``{Hadronization in the SU(3)
  Nambu-Jona-Lasinio model},''
  \href{http://dx.doi.org/10.1103/PhysRevC.53.410}{{\em Phys. Rev.} {\bf C53}
  (1996)  410--429},
\href{http://arxiv.org/abs/hep-ph/9506436}{{\tt arXiv:hep-ph/9506436
  [hep-ph]}}.
%%CITATION = HEP-PH/9506436;%%.

\bibitem{Tsue:2012jz}
Y.~Tsue, J.~da~Providencia, C.~Providencia, and M.~Yamamura, ``{Effective
  Potential Approach to Quark Ferromagnetization in High Density Quark
  Matter},'' \href{http://dx.doi.org/10.1143/PTP.128.507}{{\em Prog. Theor.
  Phys.} {\bf 128} (2012)  507--522},
\href{http://arxiv.org/abs/1205.2409}{{\tt arXiv:1205.2409 [nucl-th]}}.
%%CITATION = ARXIV:1205.2409;%%.

\bibitem{Tatsumi2011}
T.~Maruyama, E.~Nakano, and T.~Tatsumi, ``Horizons in world physics (nova
  science, ny, 2011),vol.276, chap.7,''.

\bibitem{Menezes:2008qt}
D.~P. Menezes, M.~Benghi~Pinto, S.~S. Avancini, A.~Perez~Martinez, and
  C.~Providencia, ``{Quark matter under strong magnetic fields in the
  Nambu-Jona-Lasinio Model},''
  \href{http://dx.doi.org/10.1103/PhysRevC.79.035807}{{\em Phys. Rev.} {\bf
  C79} (2009)  035807},
\href{http://arxiv.org/abs/0811.3361}{{\tt arXiv:0811.3361 [nucl-th]}}.
%%CITATION = ARXIV:0811.3361;%%.

\bibitem{Chatterjee:2011ry}
B.~Chatterjee, H.~Mishra, and A.~Mishra, ``{Vacuum structure and chiral
  symmetry breaking in strong magnetic fields for hot and dense quark
  matter},'' \href{http://dx.doi.org/10.1103/PhysRevD.84.014016}{{\em Phys.
  Rev.} {\bf D84} (2011)  014016},
\href{http://arxiv.org/abs/1101.0498}{{\tt arXiv:1101.0498 [hep-ph]}}.
%%CITATION = ARXIV:1101.0498;%%.

\bibitem{Mandal:2009uk}
T.~Mandal, P.~Jaikumar, and S.~Digal, ``{Chiral and Diquark condensates at
  large magnetic field in two-flavor superconducting quark matter},''
\href{http://arxiv.org/abs/0912.1413}{{\tt arXiv:0912.1413 [nucl-th]}}.
%%CITATION = ARXIV:0912.1413;%%.

\bibitem{Mandal:2012fq}
T.~Mandal and P.~Jaikumar, ``{Neutrality of a magnetized two-flavor quark
  superconductor},'' \href{http://dx.doi.org/10.1103/PhysRevC.87.045208}{{\em
  Phys. Rev.} {\bf C87} (2013)  045208},
\href{http://arxiv.org/abs/1209.2432}{{\tt arXiv:1209.2432 [nucl-th]}}.
%%CITATION = ARXIV:1209.2432;%%.

\bibitem{Mandal:2016dzg}
T.~Mandal and P.~Jaikumar, ``{Effect of temperature and magnetic field on
  two-flavor superconducting quark matter},''
  \href{http://dx.doi.org/10.1103/PhysRevD.94.074016}{{\em Phys. Rev.} {\bf
  D94} (2016) no.~7, 074016},
\href{http://arxiv.org/abs/1608.00882}{{\tt arXiv:1608.00882 [hep-ph]}}.
%%CITATION = ARXIV:1608.00882;%%.

\bibitem{Coppola:2017edn}
M.~Coppola, P.~Allen, A.~G. Grunfeld, and N.~N. Scoccola, ``{Magnetized color
  superconducting quark matter under compact star conditions: Phase structure
  within the SU(2)f NJL model},''
  \href{http://dx.doi.org/10.1103/PhysRevD.96.056013}{{\em Phys. Rev.} {\bf
  D96} (2017) no.~5, 056013},
\href{http://arxiv.org/abs/1707.03795}{{\tt arXiv:1707.03795 [hep-ph]}}.
%%CITATION = ARXIV:1707.03795;%%.

\bibitem{Khvorostukhin:2012kw}
A.~Khvorostukhin, V.~Toneev, and D.~Voskresensky, ``{Remarks concerning bulk
  viscosity of hadron matter in relaxation time ansatz},''
  \href{http://dx.doi.org/10.1016/j.nuclphysa.2013.07.008}{{\em Nucl. Phys. A}
  {\bf 915} (2013)  158--169}.

\bibitem{Zhuang:1995uf}
P.~Zhuang, J.~Hufner, S.~Klevansky, and L.~Neise, ``{Transport properties of a
  quark plasma and critical scattering at the chiral phase transition},''
  \href{http://dx.doi.org/10.1103/PhysRevD.51.3728}{{\em Phys. Rev. D} {\bf 51}
  (1995)  3728--3738}.

\bibitem{Gavin:1985ph}
S.~Gavin, ``{TRANSPORT COEFFICIENTS IN ULTRARELATIVISTIC HEAVY ION
  COLLISIONS},'' \href{http://dx.doi.org/10.1016/0375-9474(85)90190-3}{{\em
  Nucl. Phys. A} {\bf 435} (1985)  826--843}.

\bibitem{Kadam:2017iaz}
G.~P. Kadam, H.~Mishra, and L.~Thakur, ``{Electrical and thermal conductivities
  of hot and dense hadronic matter},''
  \href{http://dx.doi.org/10.1103/PhysRevD.98.114001}{{\em Phys. Rev.} {\bf
  D98} (2018) no.~11, 114001},
\href{http://arxiv.org/abs/1712.03805}{{\tt arXiv:1712.03805 [hep-ph]}}.
%%CITATION = ARXIV:1712.03805;%%.

\bibitem{Zhou:2020}
L.-Y. Zhou, Q.~Zheng, L.-H. Bao, and W.-J. Liang, ``{Bipolar Thermoelectrical
  Transport of SnSe Nanoplate in Low Temperature},''
  \href{http://dx.doi.org/10.1088/0256-307x/37/1/017301}{{\em Chinese Physics
  Letters} {\bf 37} (2020)  017301}.

\bibitem{Buballa:2003qv}
M.~Buballa, ``{NJL model analysis of quark matter at large density},''
  \href{http://dx.doi.org/10.1016/j.physrep.2004.11.004}{{\em Phys.Rept.} {\bf
  407} (2005)  205--376}.

\bibitem{PhysRevD.81.045015}
X.-G. Huang, M.~Huang, D.~H. Rischke, and A.~Sedrakian, ``Anisotropic
  hydrodynamics, bulk viscosities, and $r$-modes of strange quark stars with
  strong magnetic fields,'' {\em Phys. Rev. D} {\bf 81} (2010)  045015.

\bibitem{Liu:2020dxg}
S.~Y. Liu and Y.~Yin, ``{Baryonic spin Hall effect in heavy ion collisions},''
  \href{http://arxiv.org/abs/2006.12421}{{\tt arXiv:2006.12421 [nucl-th]}}.

\end{thebibliography}\endgroup
\end{document}